\documentclass[fleqn,usenatbib]{mnras}

\usepackage{newtxtext,newtxmath}
\usepackage{xcolor}

\usepackage[T1]{fontenc}

\usepackage[normalem]{ulem}

\DeclareRobustCommand{\VAN}[3]{#2}
\let\VANthebibliography\thebibliography
\def\thebibliography{\DeclareRobustCommand{\VAN}[3]{##3}\VANthebibliography}

\usepackage{graphicx}	
\usepackage{amsmath}	

\usepackage{hyperref}
\hypersetup{colorlinks,allcolors=blue}

\title[Accurate halo elongation from weak-lensing]{Accurate dark matter halo elongation from weak-lensing stacking analysis}

\author[Gonzalez et al.]{
Elizabeth J. Gonzalez,$^{1,2}$\thanks{E-mail: ejgonzalez@unc.edu.ar}
Kai Hoffmann,$^{3,4}$
Enrique Gazta\~{n}aga,$^{3,5}$
Diego R. García Lambas,$^{1,2}$
\newauthor
Pablo Fosalba,$^{3,5}$
Martin Crocce,$^{3,5}$
Francisco J. Castander,$^{3,5}$
and 
Martín Makler$^{6,7}$
\\
$^{1}$ Instituto de Astronom\'{\i}a Te\'orica y Experimental (IATE-CONICET),
 Laprida 854, X5000BGR, C\'ordoba, Argentina.\\
$^{2}$ Observatorio Astron\'omico de C\'ordoba, Universidad Nacional de C\'ordoba, Laprida 854, X5000BGR, C\'ordoba, Argentina.\\
$^{3}$Institute of Space Sciences (ICE, CSIC), Campus UAB, Carrer de Can Magrans, s/n, 08193 Bellaterra (Barcelona), Spain\\
$^{4}$ Institute for Computational Science, University of Zurich, Winterthurerstr. 190, 8057 Zürich, Switzerland\\
$^{5}$ Institut d’Estudis Espacials de Catalunya (IEEC), Barcelona, Spain \\
$^{6}$ International Center for Advanced Studies \& Instituto de Ciencias F\'isicas,  ECyT-UNSAM \& CONICET, 1650, Buenos Aires, Argentina\\
$^{7}$ Centro Brasileiro de Pesquisas F\'isicas, Rua Dr. Xavier Sigaud 150, CEP 22290-180, Rio de Janeiro, RJ, Brazil\\
}

\date{Accepted XXX. Received YYY; in original form ZZZ}

\pubyear{2015}

\begin{document}
\label{firstpage}
\pagerange{\pageref{firstpage}--\pageref{lastpage}}
\maketitle

\begin{abstract}
Shape estimates that quantify the halo anisotropic mass distribution are valuable parameters that provide information on their assembly process and evolution. Measurements of the mean shapes 
for a sample of cluster-sized halos can be used to test halo formation scenarios, as well as improving the modelling of potential biases in constraining cosmological parameters using these systems. In this work, we test the recovery of halo cluster shapes and masses applying  weak-lensing stacking techniques. To this end, we use lensing \textit{shear} and a new dark matter halo catalogue, derived from the light-cone output of the cosmological simulation MICE-GC. We perform this study by combining the lensing signals obtained for several samples of halos, selected according to their mass and redshift, taking into account the main directions of the dark-matter distributions. In the analysis we test the impact of several potential 
systematics, such as the adopted modelling, the contribution of the neighbouring mass distributions, miscentering and misalignment effects. Our results show that, when some considerations regarding the halo relaxation state are taken into account, the lensing semi-axis ratio estimates are in agreement within $5\%$ with the mean shapes of the projected dark-matter particle distribution of the stacked halos. The  methodology presented provides a useful tool to derive reliable shapes of galaxy clusters and to contrast them with those expected from numerical simulations. Furthermore, our proposed modelling, which takes into account the contribution of neighbouring halos, allows one to constrain the elongation of the surrounding mass distribution. 
\end{abstract}




\section{Introduction}

The observed large scale structure of the Universe, mainly traced by the galaxies, galaxy groups and clusters, constitutes a filamentary network. This complex structure is, according to the $\Lambda$CDM paradigm, the result of the gravitational collapse of matter driven by the initial density fluctuations. The observed luminous tracers are expected to be embedded in extended structures called halos, mostly composed by dark-matter. These halos grow in a hierarchical process as the mass is accreted in clumps mainly along the filaments, thus in preferential directions. Therefore, they are not expected to have smooth spherically symmetric mass distributions, but to show a lumpy distribution and triaxial shapes, which are 
approximately elliptical in projection. This scenario is validated by numerical simulations \citep[e.g.,][]{Dubinski1991,Warren1992,Cole1996,Jing2002,Bailin2005,Paz2006,Munoz2011,Velliscig2015,Vega-Ferrero2017} as well as by observational studies \citep[e.g.,][]{Oguri2011,Oguri2012,Lau2013,Chiu2018,Okabe2020} that obtain a clear departure of the halo shapes from spherical symmetry. 

Although the mass density distribution of halos is complex, their density profile can be simply modelled using a few parameters 
such as 
the concentration and the total mass enclosed within a defined radius \citep{Navarro97,Einasto1989,Takada2003a,Takada2003b,Baltz2009,Retana-Montenegro2012}. It has been shown that proposed theoretical profiles are efficient in modelling the density distributions within many orders of mass and along wide redshift ranges \citep{Wang2020}. 
The near universality of the mass distributions of halos places them as suitable cosmological tests since they are expected to follow well defined mass-concentration relations \citep{Neto2007,Zhao2009,Prada2012,Giocoli2012}. Moreover, the number of halos of a given mass 
is highly sensitive to some
cosmological parameters, which sets the basis for cluster cosmology. The strategy relies on the calibration of 
an observable, such as the cluster richness, that can be linked with the halo mass. However, this is susceptible to systematic effects and the halo triaxiality is identified as one of the main sources of bias introduced in the mass estimates \citep{Osato2018,McClintock2019,Zhang2022}. 

Besides the interest of studying halo shapes as potential sources of biases, characterising the shapes of the halos can also bring useful information since 
they are related to the halo formation history. There is a well-known 
trend between the halo mass, redshift and sphericity, where higher mass halos are 
on average
more elongated, since they are formed later and, therefore, 
are more affected by the last merger event. The same is expected for halos located at higher redshifts \citep{Allgood2006,Despali2017,Munoz2011}. 
Although the observed mass-shape relation has a large intrinsic scatter regarding to the different formation histories and environmental effects, it can be well constrained from simulations \citep{Munoz2011,Allgood2006,Schneider2012,Bonamigo2015,Vega-Ferrero2017}. Therefore, the observational study of the mass-shape relation can be used to test the halo formation scenario. 

Cluster halo shape determinations can also provide valuable information for the intrinsic alignments of these halos, helping to understand the alignment mechanism in general \citep{Uitert2017b}. Usually, galaxy members are used in order to measure the shape-density correlation of galaxy cluster halos \citep{Vedder2020}.
Providing more accurate halo shape measurements can be useful to improve this technique given that it can be useful to achieve more precision in the determination of the main semi-axis halo direction. Moreover, 
the cluster halo shapes can be related to the nature of the dark-matter particle \citep{Brinckmann2017,Robertson2019}. For example,  in simulations that include self-interacting dark-matter, 
halos are predicted to show a rounder particle distribution. Thus, halo shape determinations can be used to constrain the dark-matter particle cross section.

In spite of the potential applications 
of constraining halo cluster shapes, observational measurements remain as 
a challenging task. 
Gravitational lensing techniques provide a useful strategy to estimate the projected shapes of galaxy clusters. The lensing signal is related to the surface mass distribution and thus, provides information regarding the total matter content 
of these systems \citep{Diego2007,Oguri2010,Umetsu2018,Jauzac2018,Okabe2020}. In particular, weak-lensing stacking techniques enable measurements of the mean mass distribution on a wide range of distances from the cluster centres for a
combined sample of clusters, by artificially increasing the lensing signal. If these techniques are applied in a way that the combination takes into account the main orientation of the mass distribution, it allows one to measure the mean projected aligned component of the elongation, which is related to the mean surface density shape. This procedure has been successfully applied to measure the shapes of galaxy clusters \citep{Evans2009,Oguri2010,Clampitt2016,Uitert2017,shin2018,Gonzalez2021a}. 

Nevertheless, in order to relate these measurements with the predictions from simulations, it is important to take into account the potential biases that can 
can arise from this procedure.
In particular one of the main sources of biases is introduced when aligning the sample of clusters in order to perform the staking techniques. Since the main orientation of the dark-matter distribution is not known in advance, observable tracers of this orientation, such as the position of the galaxy members, are usually employed.

Given that these tracers might not be well-aligned with the underlying total mass distribution, this results in an underestimation of the surface mass elongation. In order to account for this misalignment, the lensing estimator can be boosted by including a factor, known as the dilution \citep{Clampitt2016,shin2018}. However, some assumptions need to be done to derive this dilution factor \citep{Gonzalez2021}. Other potential biases that have not been accounted for in these studies are related to the effect of miscentering, the contribution of the neighbouring mass distribution, the adopted radial range to perform the analysis and the model adopted to fit the density profiles. 

In this work we intend to conciliate the shapes derived through weak-lensing stacking techniques with the halo shapes determined according to the dark matter particle distribution. In view of the upcoming and ongoing available large weak-lensing data-sets, such as the Dark Energy Survey \citep[DES,][]{DES2005,Flaugher2005}, the Legacy Survey of Space and Time (LSST) at the NSF’s Vera Rubin Observatory \citep{Ivezic2019} and Euclid \citep{Laureijs2011}, the main goal of this study is to provide and validate a useful observational strategy to determine accurate galaxy cluster shapes that can be properly linked to the predictions provided by numerical simulations. 

To this end we use a dark-matter only simulation that includes lensing estimates which can be used to mimic the observational strategy and to properly consider the potential biases introduced in the analysis. The paper is organised as follows: we describe the halo sample used for the analysis in Sec. \ref{sec:data} and detail the adopted lensing modelling in Sec. \ref{sec:model}. In Sec. \ref{sec:comparison} we compare the estimated lensing projected shapes with the mean shapes derived according to the dark-matter particle distribution. We evaluate the potential biases introduced by the modelling and the effects of misalignment and miscentering in Sec. \ref{sec:bias}. In Sec. \ref{sec:strategy} we present an observational strategy to proper recover cluster shapes and predict the accuracy of measured shapes considering the characteristics of ongoing and future wide-field observational data-sets. Finally we summarise and conclude in Sec. \ref{sec:conclusion}.

\section{Dark-matter halos analysed}
\label{sec:data}

\begin{table*} 
 \caption{Criteria chosen to select the halo sub-samples for the stacked analysis and the number of halos included in each sub-sample.}
    \centering
    \begin{tabular}{c c c c c c c }
    \hline
    \hline
Sub-sample   & $\log M_\text{FOF}$ & $z$ & $\log(\langle M_{200} \rangle)^*$ & all halos & only relaxed 
& $q > 0.5$\\
name        &  & &  &  &  &     \\
\hline
HM-Lz & $[14.0,14.5)$  &  $[0.1,0.2)$ & 14.0 & $1325$ & $760$ & $881$\\
LM-Lz & $[13.5,14.0)$  &  $[0.1,0.2)$ & 13.5 & $6090$ & $3687$ & $4547$ \\ 
HM-Mz & $[14.0,14.5)$  &  $[0.2,0.3)$ & 14.0 & $2780$ & $1529$ & $1815$ \\
LM-Mz & $[13.5,14.0)$  &  $[0.2,0.3)$ & 13.5 & $13544$ & $7847$ & $9846$ \\
HM-Hz & $[14.0,14.5)$  &  $[0.3,0.4)$ & 14.0 & $4362$ & $2262$ & $2758$ \\
LM-Hz & $[13.5,14.0)$  &  $[0.3,0.4)$ & 13.5 & $22990$ & $12690$ & $16476$ \\
\hline
\end{tabular}
\begin{flushleft}
$^*$ Logarithm of the averaged $M_{200}$ masses obtained from fitting the 3D spherical density profiles of the halos considered with an NFW model.
\end{flushleft}
    \label{tab:sampdef}
\end{table*}

This work is based on the dark-matter only Marenostrum Institut de Ciències de l’Espai Grand Challenge (MICE-GC) simulation \citep{Fosalba2015a,Fosalba2015b,Carretero2015,Crocce2015,Hoffmann2015}. This simulation contains almost 70 billion dark-matter particles with a mass resolution of $m_\text{P} = 2.93 \times 10^{10} h^{-1} M_\odot$ in a $(3 h^{-1} $Gpc$)^{3}$ comoving volume. Halos were identified using a modified version of a publicly available halo finder\footnote{\href{http://www-hpcc.astro.washington.edu/}{http://www-hpcc.astro.washington.edu/}} applying a friends-of-friends (FOF) algorithm
 with a linking leght of $b=0.2$ times the mean inter-particle distance \citep{Crocce2015}. For the analysis we use the parameters provided by the catalogues available through the CosmoHub platform\footnote{\href{https://cosmohub.pic.es/home}{https://cosmohub.pic.es/home}} \citep{Carretero2017,TALLADA2020100391}. 

The information regarding the halos used as lenses, i.e. the halos from which we intend to recover the projected shapes using lensing techniques, is obtained from the recently released `MICE halo properties' catalogue. This catalogue lists the main properties of the halos identified in MICE-GC, such as the shape parameters and their radial density distributions. Further details of the catalogue can be found in \citet{Gonzalez2022}. The fields acquired include the centre defined according to a shrinking-sphere algorithm, labelled as `SSC', from which we obtain the halo location in the sky and redshift. We choose this centre definition since it is located at the densest halo substructure \citep{Gonzalez2022}. We consider for comparison with the lensing results, the  masses and concentrations, fitted from the density profiles computed according to the radial dark-matter particle distribution. 

The halo shape parameters presented in the catalogue were computed from the eigenvalues and the eigenvectors from the standard and reduced inertia tensors, defined as
\begin{equation}
I_{ij} \equiv  \frac{\sum_{n}^{N_p}  x_{i n} \, x_{j n}}{N_p}
\end{equation}
and 
\begin{equation} \label{eq:ired}
I^{r}_{ij} \equiv   \frac{ \sum_{n}^{N_p} (x_{i n} \, x_{jn}) / r_{n}^2 }{N_p},
\end{equation}
respectively. 
These tensors are obtained considering the positions of the $N_p$ FOF linked particles from the SSC, where $x_{in}$, $x_{jn}$, are the 2D coordinates of the vector position of the $n^{th}$ particle, and  the sub-indexes, $i,j\in[0,1]$, indicate the projected main axis. Particle coordinates are computed by defining for each halo a tangential plane perpendicular to the line-of-sight vector, pointing to the halo centre.
$r_n$ is the projected radial distance from the $n^{th}$ particle to the SSC. The semi-axis modules are provided in the catalogue and computed according the square-root of the eigenvalues ($a > b$), while the eigenvectors define the main axis directions. Since the parameters obtained from the reduced inertia tensor up-weight the particles located at the inner regions, shapes obtained from this tensor definition are more sensitive to the internal mass distribution. From the standard and reduced eigenvector components, related to the semi major-axis, we can compute $\hat{\phi}$ and $\hat{\phi}_r$, respectively, for which we refer as the standard and reduced orientations. These angles characterise the semi major-axis orientation of the projected particle distribution. Projected shapes are quantified according to the standard and reduced semi-axis ratios, $q = b/a$, where $q \rightarrow 1$ describe rounder shapes. 

For the analysis we select only the halos with FOF masses\footnote{Defined as the sum of the identified linked FOF particles times $m_\text{P}$}, $M_\text{FOF}$, higher than $10^{13.5} h^{-1} M_\odot$. Thus, all the halos included in the analysis contain more than 1000 particles and are group and cluster-sized halos. We also restrict the study to a redshift range of $0.1 < z < 0.4$. The upper limit in $z$ is mainly imposed by the resolution of the \textit{shear} components from which the lensing information is acquired. Since the lensing properties available in the mock catalogue are computed using a healpix pixelization, they are affected by the adopted angular resolution to compute the lensing maps. Therefore, the angular resolution of the \textit{shear} is $0.43$ arcmin, which corresponds to $\sim 100\,h^{-1}$\,kpc at $z=0.4$. This impacts the radial profiles at larger radii than the set by the resolution since they are sensitive to the whole integrated radial distribution. This resolution affects the inner regions of the profile reducing the radial range in which the parameters of interest are constrained. 

With the mentioned restrictions we end up with a total sample of more than $50\,000$ halos. For the analysis we also select a subset of rounder ($q > 0.5$) and a subset of relaxed halos. Relaxed halos are classified taking into account the FOF centre of mass and the kinetic and gravitational self-potential energies, $K$ and $W$, respectively. We select them considering a virial ratio cut, $2K/|W| < 1.35$, and limiting the centre of mass displacement $r_{c}/r_\text{max} < 0.1$, where $r_{c}$ is the distance from the FOF centre of mass to the SSC and $r_\text{max}$ is the radius of the first sphere which encloses all the FOF particles. We further split the  halos in six sub-samples taking into account two mass and three redshift bins. The adopted cuts and sub-samples are defined in Table \ref{tab:sampdef} where we also specify the number of halos included in each sub-sample. Given that the MICE-GC halo mass function and the mass dependence of halo clustering bias are in good agreement with observations \citep{Carretero2017,Pandey2020}, we expect that the number of stacked halos in this analysis to be comparable to the number of identified clusters in observable data-sets, for a complete sample of clusters within the same mass and redshift ranges considered.

\begin{figure}
    \centering
    \includegraphics[scale=0.6]{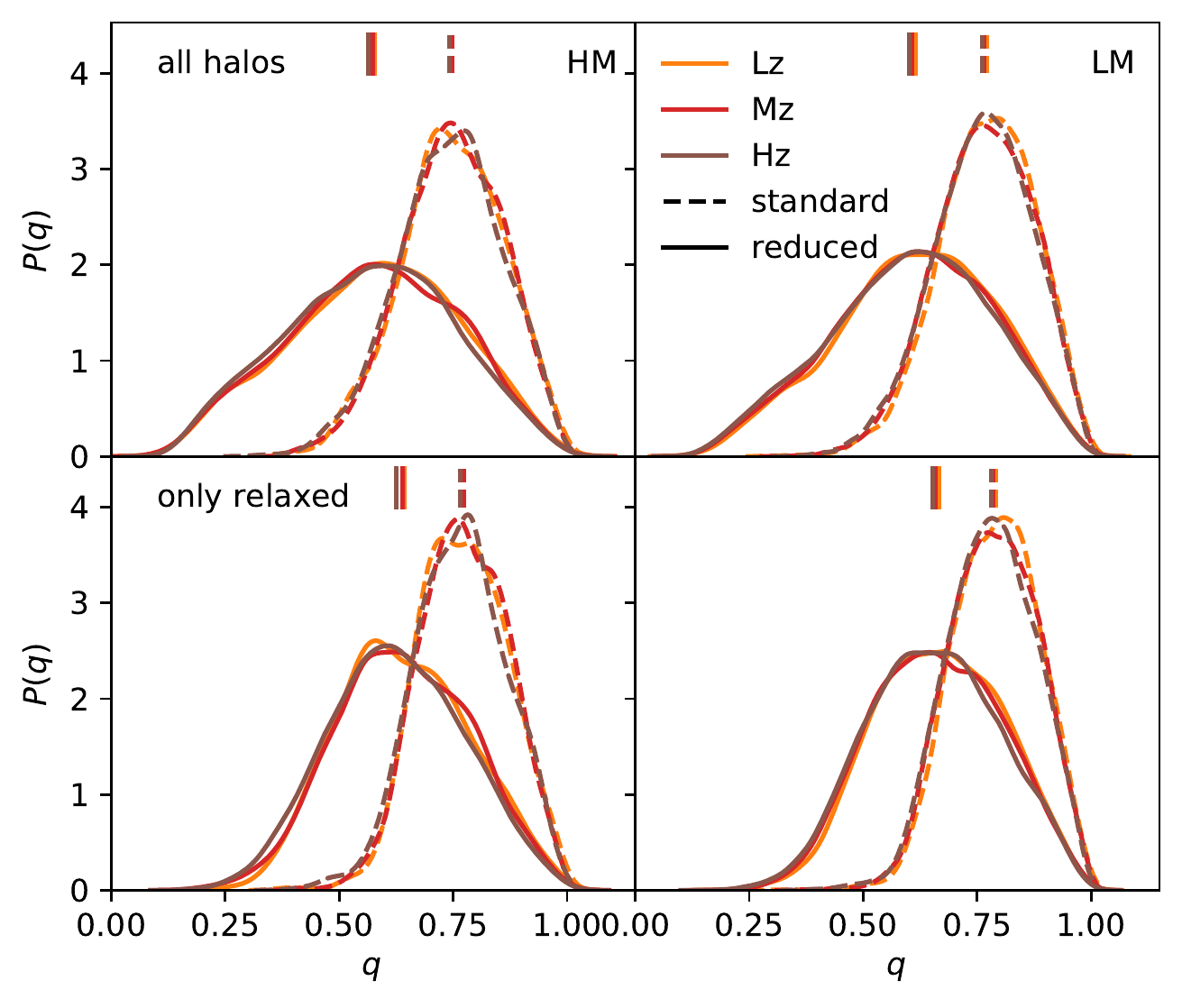}
    \caption{Probability distributions of the 2D semi-axis ratio for all (upper panel) and only relaxed (lower panel) halos included in the sub-samples described in Table \ref{tab:sampdef}. Left and right panels correspond to the higher-mass (HM: $\log M_\text{FOF} \in [14.0,14.5)$) and lower-mass (LM: $\log M_\text{FOF} \in [13.5,14.0)$) sub-samples, respectively. The colour code indicates the redshift bins of each sub-sample (Lz: $z \in [0.1,0.2)$, Mz: $z \in [0.2,0.3)$, Hz: $z \in [0.3,0.4)$). Solid and dashed lines correspond to the distribution of semi-axis ratios obtained according to the standard and reduced 
    inertia tensors, respectively. Short upper vertical lines correspond to the mean values of the distributions.}
    \label{fig:qdist}
\end{figure}

\begin{figure}
    \centering
    \includegraphics[scale=0.6]{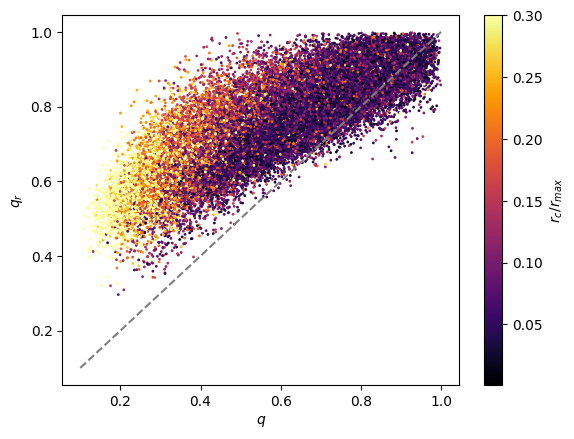}
        \includegraphics[scale=0.6]{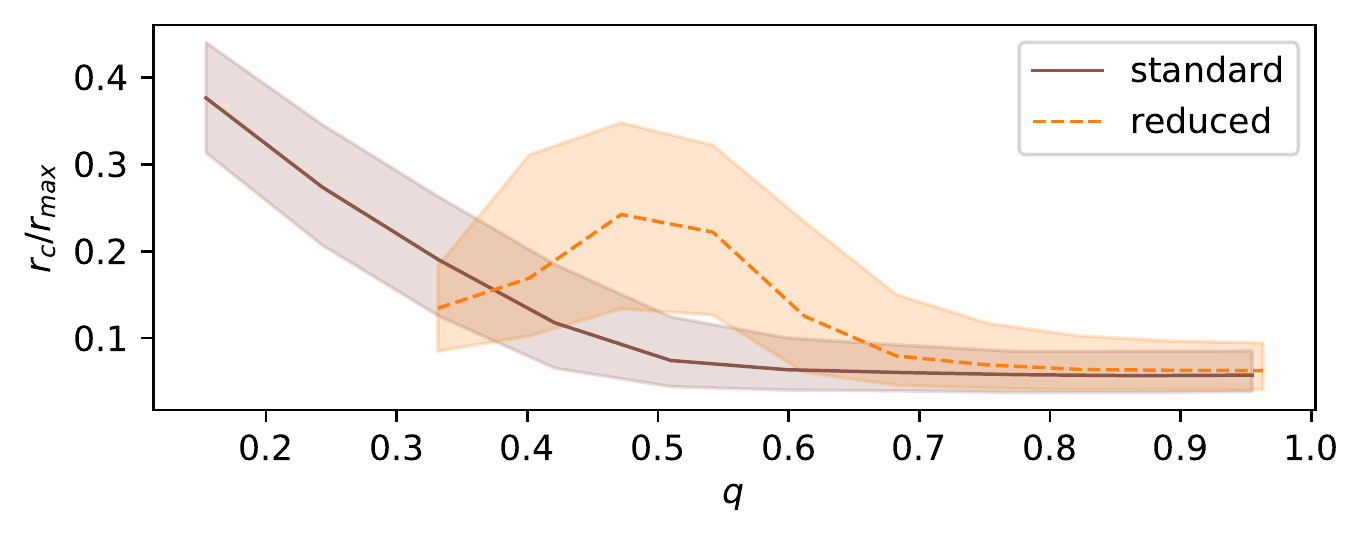}
    \caption{Upper panel: Relation between reduced ($q_r$) and standard ($q$) 2D semi-axis ratios for the total sample of 
    halos ($\log M_\text{FOF} \in [13.5,14.0)$ and $z \in [0.1,0.4)$). The colour code indicates the scaled offset between SSC and FOF centre of mass. Bottom panel: Median offsets values in bins of the semi-axis ratios. Shaded regions enclose from 25th to 75th percentiles.}
    \label{fig:qrelax}
\end{figure}

The shape parameters, $q$ and $q_r$, describe the elongation of the projected dark-matter particle distributions and are the key estimates that we intend to recover through the lensing analysis. In Fig. \ref{fig:qdist} we show the distributions of these parameters for the six defined halo sub-samples and only relaxed halos in the sub-samples. It can be noticed that reduced semi-axis ratio distributions tend to higher values describing on average rounder shapes. There is a slight variation of the mean values with redshift and mass mainly for the standard parameter (about $3\%$ and $5\%$ for the extreme redshift and mass bins, respectively), where higher mass and redshift samples tend to show lower $q$ values as expected from the halo formation scenario. When selecting only relaxed halos, the standard semi-axis ratio distributions show larger differences from the total sample of halos and shapes tend to be rounder. We also show in Fig. \ref{fig:qrelax} the relation between reduced and standard semi-axis ratios considering the scaled offset between the two halo centre definitions. As it can be noticed, halos with lower values of the standard semi-axis ratio ($q < 0.5$), which describe a highly elongated dark-matter particle distribution, include systems with larger median offsets. Therefore, these elongated halos have a main density component significantly displaced from the FOF centre, which can be indicating a large degree of substructure or that the system is undergoing a merger. We further discuss the inclusion of these halos in the analysis in Sec. \ref{sec:strategy}.

\section{weak-lensing analysis and modelling}
\label{sec:model}

The gravitational lensing effect is a physical phenomenon that distorts the shapes of extended luminous sources located behind a matter distribution named as the lens, such as a galaxy cluster. The 
shape distortion can be quantified through the \textit{shear} ($\gamma$), which is a complex quantity related to the measured ellipticity components of the source. These quantities encode information about the lens surface density distribution. In observational studies, the \textit{shear} can be estimated by measuring the ellipticities of the galaxies located behind the clusters, $e$, and then, they can be related to the projected mass distribution of these galaxy systems. To relate both quantities, $e$ and $\gamma$, several biases and systematics that are quantified, must be taken into account in the analysis. 

One of the main sources of errors comes from the fact that the observed background galaxies have intrinsic ellipticity components, $e_s$, besides those induced by the lensing effect. 
The observed source ellipticity will be, in the weak lensing regime, $e = \gamma + e_s$. To isolate the lensing contribution a statistical methodology that combines the measured ellipticity of many sources can be applied. The observed term that is related to the source's intrinsic ellipticity is averaged out and only the term related to the lensing effect is left, i.e. $\langle \epsilon \rangle = \langle \gamma \rangle$. This strategy assumes that the intrinsic orientations of background galaxies are random. Although it is well known that galaxy shapes are not randomly oriented in the sky  \citep[e.g.][]{Joachimi2015,Kiessling2015,Kirk2015,Troxel2015}, this assumption is quite accurate considering that background galaxies are selected over a wide redshift range. However, this assumption can be tested in future works making use of MICE-IA catalogue \citep{Hoffmann2022}.

The precision of the estimated \textit{shear} is related to the number of sources combined and can be optimised by adopting stacking techniques, which artificially increase the number density of sources. This procedure combines the observed signal of many lenses, thus providing mean information of the surface density distribution of the combined lenses. Besides the advantage of enhancing the lensing signal, it also reduces the intrinsic scatter introduced by the differences in the halos formation histories and smooths the density distribution by blurring the effect of substructure \citep[e.g.][]{Simet2017, Pereira2018,McClintock2019}.

The application of stacking techniques can also provide information about the mean elongation of the halos if the main orientation of the surface mass is taken into account when combining the clusters. The measured profiles obtained from the \textit{shear} components can be modelled to obtain the projected semi-axis ratio of the surface mass distribution. In this section we detail the modelling adopted to obtain this quantity, the \textit{shear} profile computation and the details of the fitting procedure. 

\subsection{Anisotropic lens model}
\label{subsec:quadru}

The lensing effect produced by an elliptical surface mass density can be obtained
by adopting a surface distribution with confocal elliptical isodensity
contours, $\Sigma(R)$, where $R$ is the 
elliptical radial coordinate, $R^2 = r^2(q \cos^2(\theta) + \sin^2(\theta)/q)$ with a semi-axis ratio $q \leq 1$ \citep{Uitert2017}.  This distribution can be approximated by considering a multipole expansion in terms of the ellipticity defined as $\epsilon:= (1-q)/(1+q)$ \citep{Schneider1991}:
\begin{equation}
\label{eq:Smyq}
\Sigma(R) = \Sigma(r,\theta) := \Sigma_0(r) + \epsilon \Sigma_2(r) \cos(2\theta),
\end{equation}
where $\theta$ is the position angle 
relative to the major semi-axis of the surface density distribution. $\Sigma_0$ and $\Sigma_2$ are the monopole and quadrupole components, respectively. $\Sigma_0$ is related to the axis-symmetrical mass distribution, which will be discussed in the next subsection,  while the quadrupole component is defined in terms of the monopole as $\Sigma_2 = -r d(\Sigma_0(r))/dr$. In this series approximation we neglect the higher order terms in $\epsilon$. 

The tangential and cross \textit{shear} components can be obtained from the deflection potential corresponding to the defined mass distribution and can also be decomposed into the monopole and quadrupole contributions:
\begin{align} \label{eq:gamma}
&    \gamma_{\rm{t}} (r,\theta) = \gamma_{\rm{t},0}(r) + \epsilon \gamma_{\rm{t},2}(r) \cos(2\theta),\\
&    \gamma_\times (r,\theta) = \epsilon \gamma_{\times,2}(r) \sin(2\theta). \nonumber
\end{align}

These \textit{shear} components are related with the surface density distribution through \citep{Uitert2017}:
\begin{align} \label{eq:gcomponents}
& \Sigma_{\rm crit} \,  \gamma_{\rm{t},0}(r)  = \frac{2}{r^2} \int^r_0 r^\prime \Sigma_0(r^\prime) dr^\prime - \Sigma_0(r),\\
& \Sigma_{\rm crit} \, \gamma_{\rm{t},2}(r) = -\frac{6 \psi_2(r)}{r^2} - 2\Sigma_0(r) - \Sigma_2(r) \nonumber, \\
& \Sigma_{\rm crit} \, \gamma_{\times,2}(r) = -\frac{6 \psi_2(r)}{r^2} - 4\Sigma_0(r),  \nonumber
\end{align}
where $\psi_2(r)$ is the quadrupole component of the lensing potential and is obtained as:
\begin{equation}\label{eq:psi2}
    \psi_2(r) = -\frac{2}{r^2} \int_0^r r^{\prime 3} \Sigma_0(r^\prime) dr^{\prime}.
\end{equation}
\textit{Shear} components are multiplied by the critical density which contains all the geometrical information about the observer-lens-source configuration and is defined as:
\begin{equation} \label{eq:sig_crit}
\Sigma_{\rm{crit}} = \dfrac{c^{2}}{4 \pi G} \dfrac{D_\text{OS}}{D_\text{OL} D_\text{LS}},
\end{equation}
where $D_\text{OL}$, $D_\text{OS}$ and $D_\text{LS}$ are  the angular diameter distances from the observer to the lens, from the observer to the source and from the lens to the source, respectively. In this way we can combine the observed signal at different source and lens redshifts, obtaining distance independent quantities.
The tangential component of the monopole is the usual density contrast
 $\Delta \Sigma$,
\begin{equation} \label{eq:DSigma}
\Delta \Sigma(r) =  \Sigma_{\rm crit} \gamma_{\rm{t},0}(r) = \frac{1}{2\pi} \int_0^{2\pi} \Sigma_{\rm crit}\, \gamma_{\rm{t}} (r,\theta) d\theta   
\end{equation}
which is the only term observed in the case of an axis-symmetric mass distribution.

If we average the tangential and cross components projected according to the position angle $\theta$ in annular bins, we  can isolate the quadrupole components scaled according to the ellipticity:
\begin{align} \label{eq:gproj1}
& \Gamma_{\rm{T}}(r) := \epsilon \Sigma_{\rm crit} \gamma_{\rm{t},2}(r) = \frac{1}{\pi} \int_0^{2\pi} \Sigma_{\rm crit} \gamma_{\rm{t}} (r,\theta) \cos(2\theta) d\theta, \\
\label{eq:gproj2}
& \Gamma_{\times}(r) := \epsilon \Sigma_{\rm crit} \gamma_{\times,2}(r) = \frac{1}{\pi} \int_0^{2\pi} \Sigma_{\rm crit} \gamma_\times (r,\theta) \sin(2\theta) d\theta,
\end{align}
where we defined the distance independent quantities related with the quadrupole, $\Gamma_{\rm{T}}$ and $\Gamma_\times$.

\subsection{Surface distribution modelling}
\label{subsec:mono}

Usually, in observational studies of galaxy clusters that consider a stacked analysis \citep[e.g.][]{Johnston2007,Simet2017,Luo2018,Pereira2020}, the tangential \textit{shear} profile is modelled by considering a sum of terms that mainly takes into account: (1) a component related to the contribution of the brightest cluster galaxy member (BCG) which is often adopted as the cluster centre; (2) the main halo component related to the contribution of the cluster host halo and which is assumed to be well centred at the BCG; (3) a component that takes into account a population of clusters that are not properly centred at the BCG which are mainly related to merging systems and, finally, (4) a component related to the mass distribution of the neighbouring halos usually referred to as the 2-halo term. In our case, we model the radial surface density distribution by taking into account the main halo host and the neighbouring mass contributions (components 2 and 4). Thus, we neglect the contribution of the BCG subhalo and of the miscentering term. The impact of neglecting the miscentering in the quadrupole components will be discussed in Sec. \ref{sec:bias}, while the contribution of the central subhalo is avoided by fitting the profile beyond $350\,h^{-1}$\,kpc. 
By avoiding smaller scales we expect to reduce errors in the computation of the \textit{shear} from the resolution of the simulation.
We are in turn consistent with observational analyses, which avoid small scales to reduce systematic effects, such as obscuration, significant membership contamination, and blending \citep{McClintock2019}.

 The main halo component is modelled assuming a spherically symmetric NFW profile \citep{Navarro97}, which depends on two parameters, the radius that encloses a mean density equal to 200 times the critical density of the Universe, $R_{200}$, and a dimensionless concentration parameter, $c_{200}$. The 3D density profile is given by:
\begin{equation} \label{eq:nfw}
\rho_{1h}(r) =  \dfrac{\rho_{\rm crit} \delta_{c}}{(r/r_{s})(1+r/r_{s})^{2}},
\end{equation}
where  $r_{s}$ is the scale radius, $r_{s} = R_{200}/c_{200}$, $\rho_{\rm crit}$ is the critical density of the Universe at the mean redshift ($\langle z \rangle$) of the sample of stacked halos and
$\delta_{c}$ is the cha\-rac\-te\-ris\-tic overdensity:
\begin{equation}
\delta_{c} = \frac{200}{3} \dfrac{c_{200}^{3}}{\ln(1+c_{200})-c_{200}/(1+c_{200})}.  
\end{equation}
We compute $\langle z \rangle$ by averaging the redshifts of the clusters in the sample weighted according to the number of source galaxies considered for each halo to compute the profile.
The mass within $R_{200}$ can be obtained as \mbox{$M_{200}=200\,\rho_{\rm crit} (4/3) \pi\,R_{200}^{3}$}. 

The 2-halo term is related to the excess in the halo mass distribution introduced by the neighbouring halos and depends on the mass of the main halo component. The 3D density profile related to this term is obtained considering the halo-matter correlation function, $\xi_{hm}$, as:
\begin{equation} \label{eq:rho2h}
    \rho_{2h}(r) = \rho_{m} \xi_{hm} = \rho_{\rm crit} \Omega_m (1+z)^3 b(M_{200},\langle z \rangle) \xi_{mm}
\end{equation}
where $\rho_m$ is the mean density of the Universe ($\rho_m = \rho_{\rm crit} \Omega_m (1+z)^3$) and the halo-matter correlation function is related to the matter-matter correlation function through the halo bias \citep[$\xi_{hm}  = b(M_{200},\langle z \rangle) \xi_{mm}$,][]{Seljak2004}. We set the halo bias by adopting \citet{Tinker2010} model calibration.

The total surface density profile is modelled by considering an elliptical distribution for the main halo component with an elongation $\epsilon_{1h}$ plus the term introduced by the neighbouring distribution also elongated and characterised by the aligned ellipticity component, $\epsilon_{2h}$. Thus the total projected surface density profiles are modelled as:
\begin{equation} \label{eq:smodel}
    \Sigma(R) \sim \Sigma_{1h}(r) + \epsilon_{1h} \Sigma^{\prime}_{1h}(r) \cos(2\theta) + \Sigma_{2h}(r) + \epsilon_{2h} \Sigma^{\prime}_{2h}(r) \cos(2\theta)
\end{equation}
where $\Sigma_{1h}$ corresponds to the projected NFW centred halo profile and $\Sigma_{2h}$ is the projected density of the neighbouring mass, derived according to Equations \ref{eq:nfw} and \ref{eq:rho2h}, respectively. The quadrupoles are related to each monopole component as $\Sigma^{\prime}_{1h} = -r d(\Sigma_{1h}(r))/dr$ and $\Sigma^{\prime}_{2h} = -r d(\Sigma_{2h}(r))/dr$. Finally $\theta$ is the position angle of the source with respect to the major semi-axis of the halo mass distribution as in Eq. \ref{eq:smodel}. In an analogous way, we model the \textit{shear} profiles described in Eq. \ref{eq:DSigma}, \ref{eq:gproj1} and \ref{eq:gproj2} as a sum of these surface components. Both monopoles terms, $\Sigma_{1h}$ and $\Sigma_{2h}$, are computed by using \textsc{COLOSSUS}\footnote{\href{https://bitbucket.org/bdiemer/colossus/src/master/}{https://bitbucket.org/bdiemer/colossus/src/master/}} astrophysics toolkit \citep{Diemer2018}. With this adopted model of the monopole, we are assuming that the density 
distribution of the main halo as well as for the contribution of the neighbouring masses is elongated through the whole radial range along the same semi-axis ratio. The expected misalignment between the main halo component and the neighbouring distribution will bias the $\epsilon_{2h}$ quantity to lower values. 

\subsection{Mock source galaxy sample}
\label{subsec:sources}

The weak-lensing data-set used in this work is based on MICECAT v2.0 mock catalogue. The total area of this catalogue is 1/8 of the sky with a galaxy density of $\sim 27$ galaxies/arcmin$^{-2}$. We compare the catalogue characteristics with ongoing and future observed projects in Table \ref{tab:wldataset}. As we can see, 
the source density of the mock data used in this work is comparable with the ones expected for future wide-surveys data-sets albeit with a lower total sky area.

\begin{figure}
    \centering
    \includegraphics[scale=0.6]{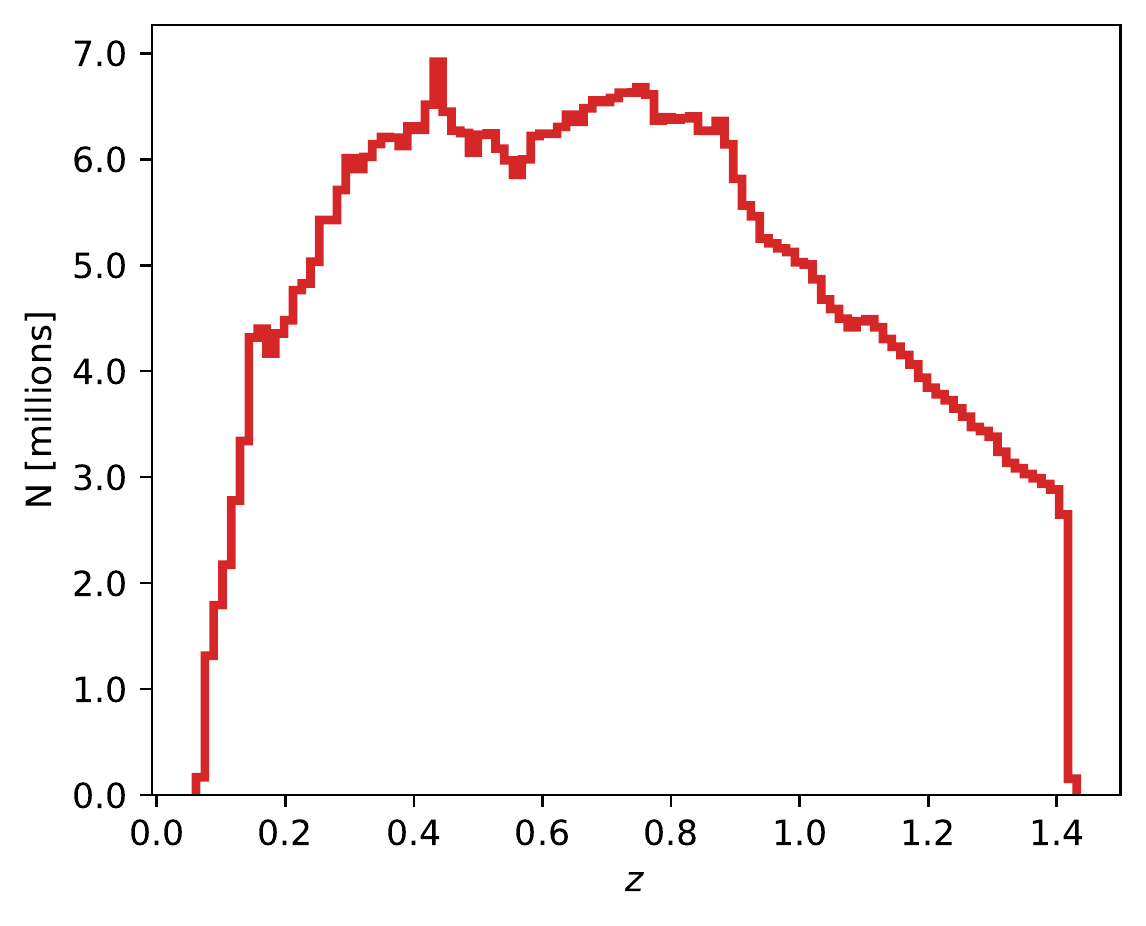}
    \caption{Redshift distribution of the source galaxies.}
    \label{fig:zdist}
\end{figure}

Source galaxies are defined as the galaxies that are behind the halos described in Sec. \ref{sec:data}, hence, affected by the lensing effect due to the gravitational potential generated by the halos. Taking into account the galaxy locations, we select from the mock catalogue the source galaxies as those with redshifts higher than the halo redshift plus $0.1$. In Fig. \ref{fig:zdist} we show the redshift distribution of the source galaxies. In average for each halo we use about 
$10^6$ ($10^{4.5}$) galaxies
to compute the stacked profiles for the lowest (highest) redshift bins considered. 

It is important to notice 
that to perform the stacking analysis we combine the \textit{shear} components, $\gamma$, without considering the galaxy intrinsic shapes, $e_s$. Therefore our estimates are not going to be affected by the noise due to the intrinsic ellipticity dispersion. In order to predict how this would impact in the shape measurements using observational data, we introduce this noise in Sec. \ref{sec:strategy}.

\begin{table} 
 \caption{Weak-lensing data-set characteristics used in this work compared with ongoing and future observational wide-field surveys.}
    \centering
    \begin{tabular}{l c c c c}
    \hline
    \hline
& MICECAT v2.0 & DES & Euclid & LSST \\
 & & & &  \\
Area & 5\,157 & 5\,000 & 15\,000 & 30\,0000 \\
$[\text{deg}^2]$ & & & &  \\
Source density  & 26.9 & 6.28 & $\sim 30$ & $\sim 30$ \\
$[\text{arcmin}^{-2}]$ &  &  &  &  \\
\hline
\end{tabular}
\begin{flushleft}
\end{flushleft}
    \label{tab:wldataset}
\end{table}

\subsection{Estimators and computed profiles}
\label{subsec:estimators}

In order to obtain the monopole and quadrupole stacked profiles, we combine the \textit{shear} components of all the sources for the halos included in each sample defined in Table \ref{tab:sampdef}. Profiles are obtained using 40 non-overlapping concentric logarithmic annuli to preserve the signal-to-noise ratio of the outer region, from $100\,h^{-1}$\,kpc up to $10\,h^{-1}$\,Mpc, using the following 
estimators:
\begin{align} 
& \widetilde{\Delta \Sigma}(r) = \frac{\sum_{j=1}^{N_L} \sum_{i=1}^{N_{S,j}} \Sigma_{{\rm crit},ij} \gamma_{{\rm t},ij}}{\sum_{j=1}^{N_L} N_{S,j} }, \label{eq:rprof} \\
& \widetilde{\Gamma}_{\rm{T}}(r) = \frac{\sum_{j=1}^{N_L} \sum_{i=1}^{N_{S,j}} \Sigma_{{\rm crit},ij} \gamma_{{\rm t},ij} \cos{2 \hat{\theta}_i}}{\sum_{j=1}^{N_L} \sum_{i=1}^{N_{S,j}}  \cos^2{2\hat{\theta}_i}}, \label{eq:profgammatan}\\
& \widetilde{\Gamma}_{\times}(r) = \frac{\sum_{j=1}^{N_L} \sum_{i=1}^{N_{S,j}} \Sigma_{{\rm crit},ij} \gamma_{{\rm \times},ij} \sin{2\hat{\theta}_i}}{\sum_{j=1}^{N_L} \sum_{i=1}^{N_{S,j}} \sin^2{2\hat{\theta}_i}},\label{eq:profgammacross}
\end{align}
where $\gamma_{{\rm t},ij}$ and $\gamma_{{\rm \times},ij}$ are the tangential and the cross \textit{shear} components for the $i-$th source of the $j-$th lens. $N_L$ is the number of halos considered as lenses for the stacking and $N_{S,j}$ the number of sources located at a projected physical distance $r \pm \delta r$ from the $j$th halo. $\Sigma_{{\rm crit},ij}$ is the critical density defined in Eq. \ref{eq:sig_crit} computed for the $i-$th source of the $j-$th lens. $\hat{\theta}_i$ is the adopted position angle of the source with respect to the main orientation of the halo surface mass distribution. This angle is computed using the major semi-axis direction according to the orientation of the projected positions of the FOF dark-matter particles, $\hat{\phi}$ and $\hat{\phi}_r$, defined in the previous section.

The associated covariance to each estimator is computed using a jacknife resampling, by splitting the total field into $K=100$ equally sized sky regions, $S_k$. The jacknife covariance matrix is then obtained as:
\begin{equation}
\label{eq:cov}
    C_{\widetilde{E}} = \frac{K-1}{K} \sum_{k=1}^K (\widetilde{E}(k) - \widetilde{E}(\cdot))^T \cdot (\widetilde{E}(k) - \widetilde{E}(\cdot))
\end{equation}
where $\widetilde{E}$ corresponds to the estimators: $\widetilde{\Delta \Sigma}(r)$, $\widetilde{\Gamma}_{\rm{T}}$ and $\widetilde{\Gamma}_{\times}$ defined in Eqs. \ref{eq:rprof}--\ref{eq:profgammacross}. $\widetilde{E}(\cdot) = (1/K) \sum_k \widetilde{E}(k)$ and $\widetilde{E}(k)$ is the estimator computed using all the combined lenses except for those in the $S_k$ region.

\subsection{Fitting procedure}
\label{subsec:fit}

Profiles are fitted within a radial range $350\,h^{-1}$\,kpc $ < r < 5\,h^{-1}$\,Mpc. The inner range is selected to avoid the numerical effects which deplete the \textit{shear} signal \citep{Oguri2011} while the external radial range is a usually adopted limit in similar observational studies \citep[e.g.][]{Pereira2018,Gonzalez2021,McClintock2019}. Due to the resolution of the \textit{shear} components that mostly affects to higher-redshift sub-samples, highly biased results were obtained for HM-Hz and LM-Hz. The depletion in the inner region tends to bias the results by fitting lower mass values than the expected. Fitted masses tend to be more in agreement with the fitted dark-matter particle density profiles when adopting a larger inner radii of $450\,h^{-1}$\,kpc. Nevertheless, for the less massive sub-sample, LM-Hz, fitted quadrupole components were poorly constrained given that the contribution of the neighbouring mass distribution is more significant at lower radii for the lower mass halos. Taking this into account we adopt inner radii to fit the profiles of $450\,h^{-1}$\,kpc and $400\,h^{-1}$\,kpc for the HM-Hz and LM-Lz, respectively.

The parameters that are fitted from the 
estimator profiles are the logarithmic mass, $\log(M_{200})$, the concentration, $c_{200}$, the average aligned semi-axis ratio of the main halo component, $\tilde{q}_{1h}$, and the one related to the neighbouring mass distribution $\tilde{q}_{2h}$. These are defined taking into account the aligned elongation components introduced in Eq. \ref{eq:smodel} ($\tilde{q}_{1h} = (1 - \epsilon_{1h})/(1 + \epsilon_{1h})$ and $\tilde{q}_{2h} = (1 - \epsilon_{2h})/(1 + \epsilon_{2h})$). 

Our fitting procedure consists of two-steps, first we obtain $\log(M_{200})$ and $c_{200}$ by fitting only the profile that corresponds to the monopole component ($\widetilde{\Delta \Sigma}(r)$). Then, taking into account the estimated $M_{200}$ and $c_{200}$, we constrain $\tilde{q}_{1h}$ and $\tilde{q}_{2h}$ by simultaneously fitting the quadrupole components, $\widetilde{\Gamma}_{\rm{T}}(r)$ and $\widetilde{\Gamma}_{\times}(r)$. The procedure adopted is mainly motivated by the fact that the signal-to-noise ratios of the quadrupole profiles are significantly lower than the one obtained for the monopole component alone. Thus, the monopole can be fitted to obtain the mass and concentration with a smaller error and then fixed to obtain the projected semi-axis ratios, optimising the parameter determination. We test the differences between the adopted two-step procedure and by simultaneously fitting the three 
profiles to obtain the four free parameters. In general, no significant differences are obtained between the fitted parameters, which are mostly in agreement within $1\%$. Therefore, we conclude that our results do not depend on the choice of the fitting methodology. 

We constrain our free parameters by using the Markov chain Monte Carlo (MCMC) method, implemented through \texttt{emcee} python package \citep{Foreman2013}, to optimise the  log-likelihood functions for the monopole profile:
\begin{equation}
\label{eq:loglmono}
\ln{\mathcal{L}}(\Delta \Sigma | M_{200},c_{200}) = \frac{1}{2} (\widetilde{\Delta \Sigma} - \Delta \Sigma) C^{-1}_{\Delta \Sigma} (\widetilde{\Delta \Sigma} - \Delta \Sigma)
\end{equation}
 and for the quadrupoles we minimise the sum of the likelihoods $\ln{\mathcal{L}}(\Gamma_{\rm{T}} | r ,\tilde{q}_{1h}, \tilde{q}_{2h}) + \ln{\mathcal{L}}(\Gamma_{\times} | r ,\tilde{q}_{1h}, \tilde{q}_{2h})$, defined as:
\begin{equation}
\ln{\mathcal{L}}(\Gamma_{\rm{T}} | \tilde{q}_{1h}, \tilde{q}_{2h}) = \frac{1}{2} D_{\Gamma_{\rm{T}}} C^{-1}_{\Gamma_{\rm{T}}} D_{\Gamma_{\rm{T}}}
\end{equation}
\begin{equation}
\ln{\mathcal{L}}(\Gamma_{\times} | \tilde{q}_{1h}, \tilde{q}_{2h}) = \frac{1}{2} D_{\Gamma_{\times}} C^{-1}_{\Gamma_{\times}} D_{\Gamma_{\times}}
\end{equation}
where $D_{\Gamma_{\rm{T}}} = \widetilde{\Gamma}_{\rm{T}} - \Gamma_{\rm{T}}$ and $D_{\Gamma_{\times}} = \widetilde{\Gamma}_{\times} - \Gamma_{\times}$. To fit the data we use 15 chains for each parameter and 250 steps, considering flat priors for the three parameters: $12.5 < \log(M_{200}/(h^{-1} M_\odot)) < 15.5$, $1 < c_{200} < 5$, $0.2 < \tilde{q}_{1h} < 0.9$ and $0.2 < \tilde{q}_{2h} < 0.9$. Our best fit parameters are obtained after discarding the first 50 steps of each chain, according to the median of the marginalised posterior distributions and errors enclose the central $64$\% of the marginalised posterior. 

\begin{figure*}
    \centering
    \includegraphics[scale=0.6]{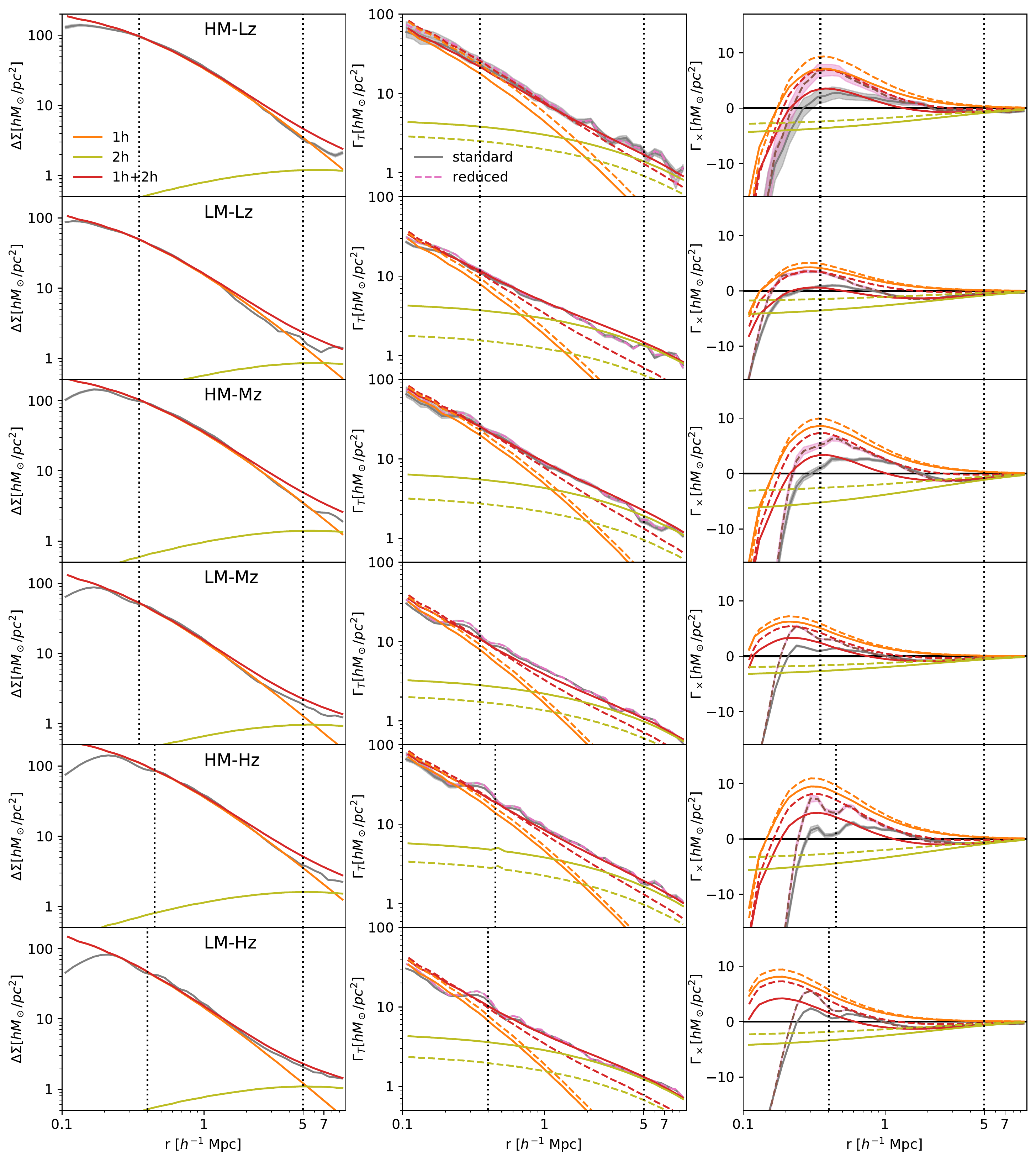}
    \caption{Stacked density contrast profiles ($\Delta \Sigma$, first column), tangential quadrupole component profiles ($\Gamma_{\text{T}}$, second column) and cross quadrupole component profiles ($\Gamma_{\times}$, third column) for the sub-samples defined in Table \ref{tab:sampdef}. Quadrupole profiles are computed considering the standard ($\hat{\phi}$, solid grey lines) and reduced orientations ($\hat{\phi}_r$, dashed pink lines). Shaded regions correspond to the square root of the diagonal components of the covariance matrix defined in Eq. \ref{eq:cov}.
    The regions enclosed within the dotted vertical lines indicate the fitted radial ranges. Together with the profiles we show the correspondent fitted models for the main halo component (orange line), the 2-halo term component (yellow line) and the sum of both fitted components (red). Solid and dashed lines for the fitted models in the quadrupole profiles 
    correspond to the profiles computed using the standard and reduced orientations, respectively.} 
    \label{fig:profile}
\end{figure*}

\section{Conciliating shape estimates}
\label{sec:comparison}

In order to compare the estimated mean halo semi-axis ratio obtained from the lensing analysis, $\tilde{q}_{1h}$, with the one derived according to the dark-matter particle distribution, we compute the monopole and quadrupole profiles for the sub-samples of halos presented in Table \ref{tab:sampdef}. We keep all the identified halos in the lensing analysis within the selected mass and redshift ranges regardless of the relaxation state. We 
discuss a different strategy in Sec. \ref{sec:strategy}. 

Lensing quadrupole profiles are obtained using Eqs. 
\ref{eq:profgammatan} and \ref{eq:profgammacross}
considering both standard and reduced halo orientations: $\hat{\phi}$ and $\hat{\phi}_r$, determined according to the particle distribution as defined in Sec. \ref{sec:data}. The distribution of $\hat{\phi} - \hat{\phi}_r$ has a standard deviation of $36^\circ$ and we expect, according to the inertia tensor definitions, that $\hat{\phi}$ to be more aligned with the external mass distribution than $\hat{\phi}_r$.
In observational studies both quantities can be related to the adopted tracer to estimate the main direction of the surface mass distribution. For example, when considering the orientations estimated according to the BCG or the distribution of the red galaxy members, we expect for them to be more related to $\hat{\phi}_r$, since these tracers follow more tightly the inner regions of the mass distribution. Otherwise, when considering bluer galaxy members to estimate the main direction we expect them to be more related to the standard orientation
\citep{vanUitert2017,Gonzalez2021,Gonzalez2022}.
This expectations can be tested later in future works that include hydrodinamical simulations. 

By fitting the profiles using the modelling described by Eq. \ref{eq:smodel}, we obtain for each sub-sample of halos two sets of semi-axis ratio estimators when applying the lensing analysis, $(\tilde{q}_{1h}(\hat{\phi}), \tilde{q}_{2h}(\hat{\phi}))$ and $(\tilde{q}_{1h}(\hat{\phi}_r), \tilde{q}_{2h}(\hat{\phi}_r))$.

\label{subsec:part}

\begin{figure}
    \centering
    \includegraphics[scale=0.6]{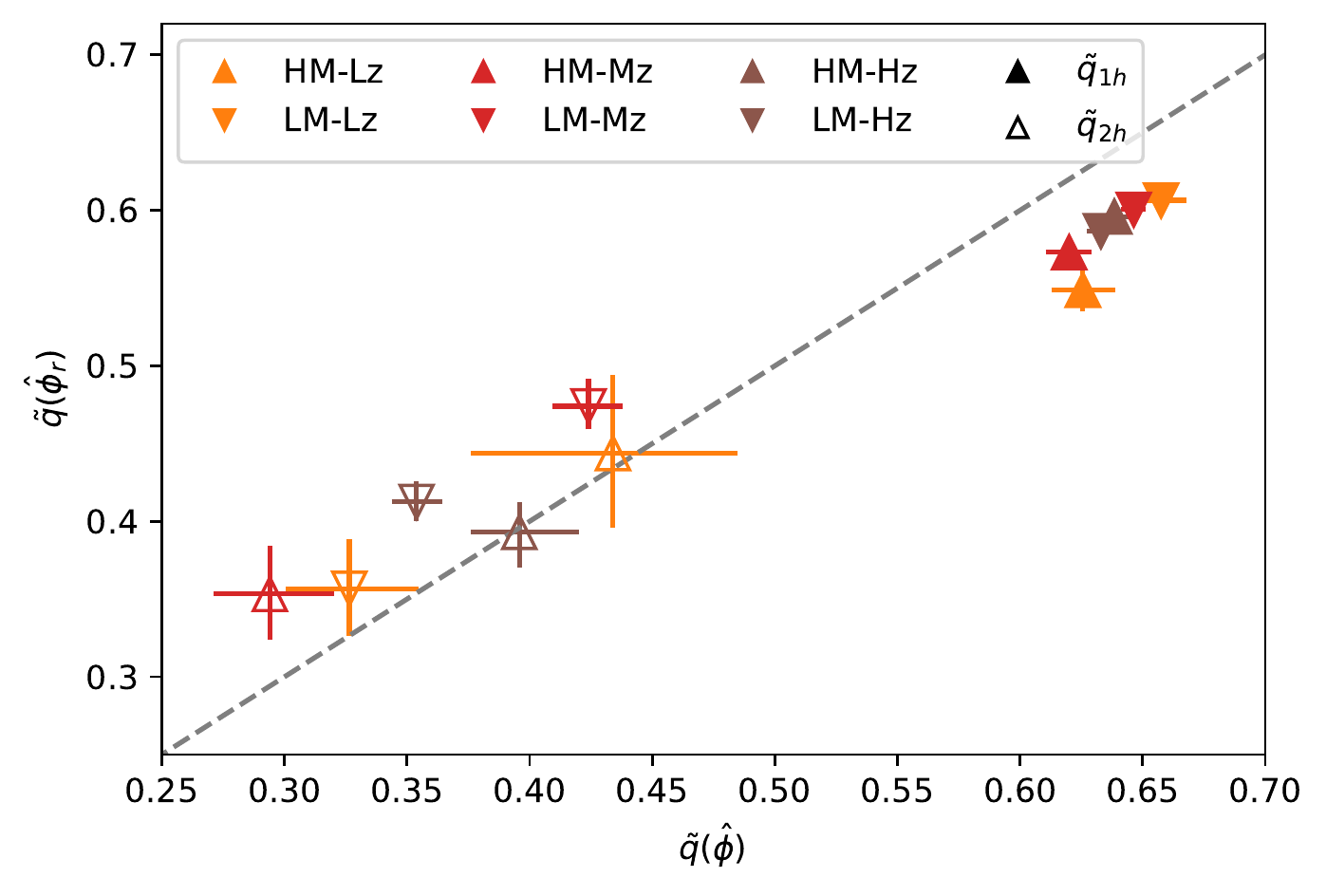}
    \caption{Fitted semi-axis ratios using the standard vs. the reduced orientations ($\phi$ and $\phi_r$, respectively) to compute the quadrupole profiles. Symbols represent each sub-sample of halos. Full and empty symbols 
    refer to the main halo component and the surrounding mass distribution semi-axis ratios, $\tilde{q}_{1h}$ and $\tilde{q}_{2h}$, respectively.}
    \label{fig:q2h}
\end{figure}

\subsection{Fitting results}
We show the fitted profiles in Fig. \ref{fig:profile}: the density contrast distribution, $\Delta \Sigma$, and the tangential and cross quadrupole components, $\Gamma_{\rm{T}}$ and $\Gamma_{\times}$, respectively. For the quadrupole components we show the ones computed using the standard (solid lines) and the reduced (dashed lines) orientations, with their fitted models. The cross component profiles show higher differences when considering reduced and standard orientations within the central regions up to $1 h^{-1}$Mpc, in comparison with the results for the tangential component. 
In the most inner regions ($\sim 350 h^{-1}$kpc) tangential and cross quadrupoles components using reduced orientations are up to $1.1$ and $1.4$ times the values computed using the standard orientation, respectively. Since the quadrupole components are only sensitive to the enclosed mass distribution, when considering reduced orientations the alignment with the mass distribution is higher in the inner regions resulting in a higher lensing signal. The differences obtained from both components can be related to a radial variation of the particle distribution elongation, which is not considered in the modelling. Previous works report that the dark-matter particle distribution of halos tend to be more elongated in the inner regions \citep{Warren1992,Jing2002,Bailin2005,Allgood2006,Despali2017}. This trend is reversed when baryons are incorporated \citep[e.g.,][]{Chua2019,Gonzalez2021}. Therefore, we expect different results in observational works.

In general, lensing mass estimates fitted from the $\widetilde{\Delta \Sigma}$ profiles are $10\%$ biased to lower values in comparison with the fitted masses from the 3D density profiles. Also, fitted concentrations are significantly biased, with the lensing fitted concentrations being lower by $10-30\%$. These biases show no significant differences with the mean mass and redshift of the stacked halos. These 
discrepancies can be related to the differences in the fitted radial ranges, projection effects and the density profile model adopted, as stated in \citet{Gonzalez2022}. We 
shall not
inspect further the possible bias introduced in measuring masses in this work. However we will discuss in the next sections how systematics in the mass measurements and concentrations can impact recovering the halo shapes. 

In Fig. \ref{fig:q2h} we compare the lensing estimates using standard and reduced directions, for both constrained semi-axis ratios,  $\tilde{q}_{1h}$ and  $\tilde{q}_{2h}$. Semi-axis ratios obtained for the main halo component are considerably higher than those fitted for the surrounding contribution, indicating that the averaged neighbouring mass distribution is well-aligned with the halo and highly elongated. When considering the reduced main directions, estimated shapes describe a more elongated main halo and a rounder neighbouring distribution than when using the standard main direction. This is expected since the parameters based on the standard inertia tensor are more sensitive to the internal regions.

\begin{figure}
    \centering
    \includegraphics[scale=0.6]{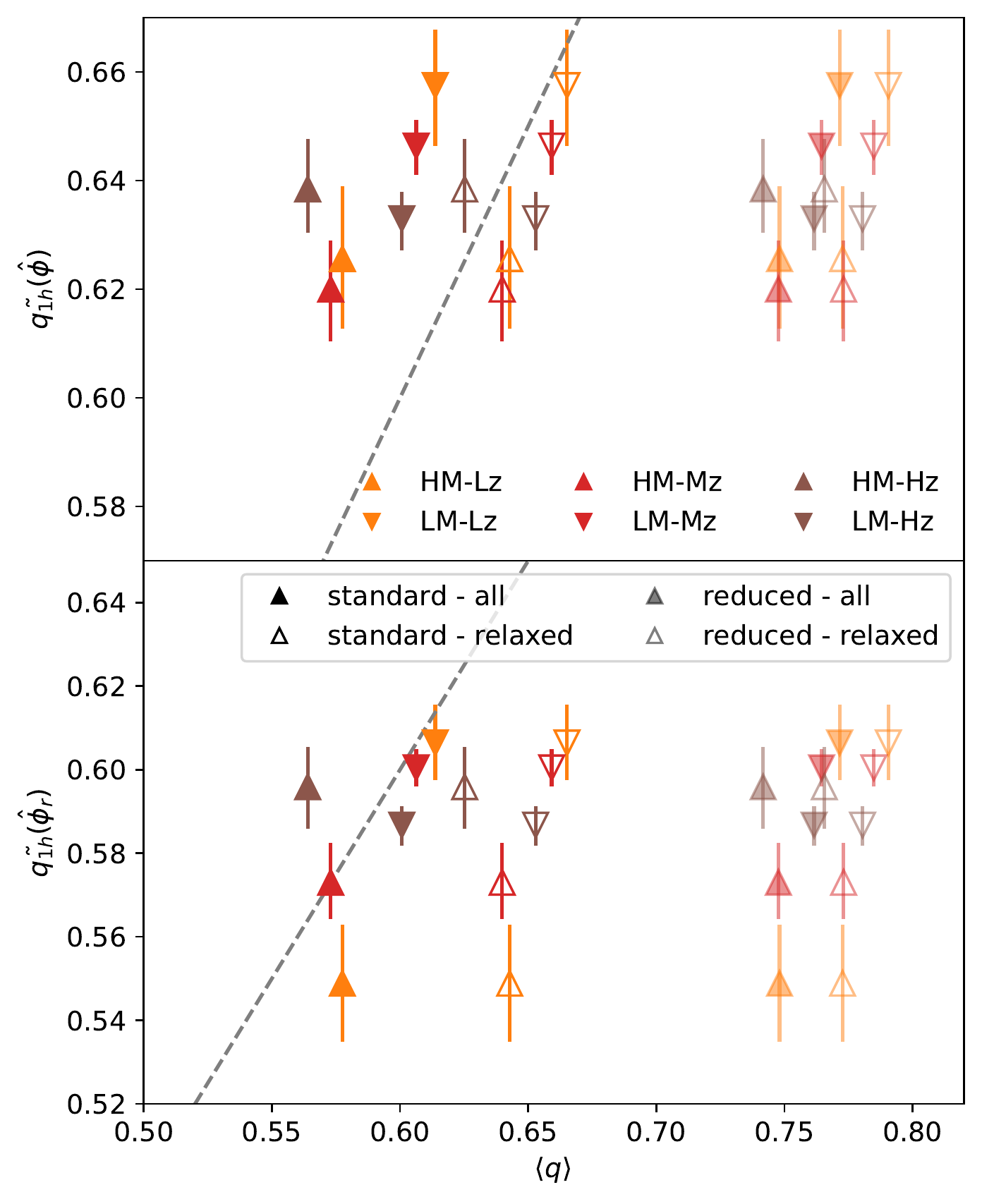}
    \caption{Lensing fitted main semi-axis ratio component ($\tilde{q}_{1h}$) computed considering the standard (upper panel) and reduced (lower panel) orientations, compared with the mean 2D semi-axis ratio derived from the standard (solid symbols) and reduced (dashed symbols) inertia tensors. Symbols and the colour code represent the sub-sample of halos defined in Table \ref{tab:sampdef}. Open symbols represent the mean values computed considering only relaxed halos. Dashed grey line corresponds to the identity function.}
    \label{fig:qcomp}
\end{figure}

\subsection{Projected dark-matter semi-axis ratio vs. lensing estimator}
\label{subsec:comp}
In Fig. \ref{fig:qcomp} we compare the fitted lensing semi-axis ratios together with the mean of the semi-axis ratios obtained from the dark-matter particle distribution, $\langle q \rangle$, using the standard and reduced inertia tensors. For the comparison we compute the mean by considering the whole sample of halos within the selected FOF mass and redshift ranges and, also, with a further constraint by including only relaxed halos. These parameters quantify the mean halo shapes, hence they are related to the fitted $\tilde{q}_{1h}$ lensing parameter. As it can be noticed, lensing estimates are in agreement with the $\langle q \rangle$ values computed using the standard inertia tensor, while the shapes estimated using the reduced inertia tensor are significantly rounder. The good general agreement between the lensing estimates and $\langle q \rangle$ values obtained according the standard tensor instead of the reduced one, is due to the fact that the weak-lensing effect is more sensitive to the external region of the halo mass distribution. 

\begin{figure*}
    \centering
    \includegraphics[scale=0.6]{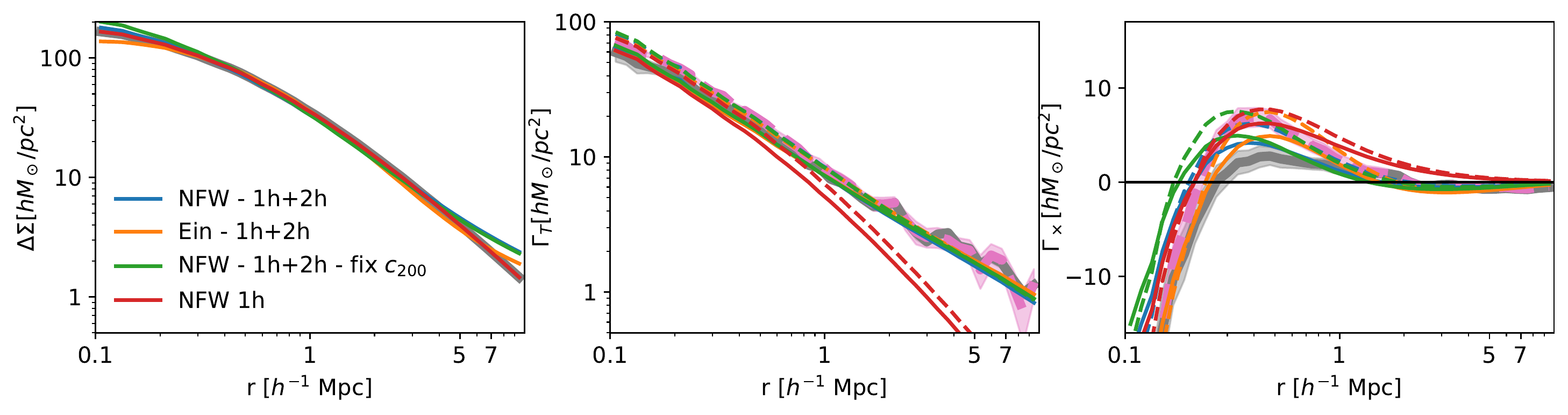}
    \caption{Density contrast, tangential and cross quadrupole profiles obtained for the stacked HM-Lz sub-sample of halos in grey, when computing the quadrupole profiles using the standard orientation and in pink using reduced orientations. Solid and dashed lines in blue, orange, green and red correspond to the indicated fitted models, for standard and reduced orientations, respectively.}
    \label{fig:promodel}
\end{figure*}

\begin{figure*}
    \centering
    \includegraphics[scale=0.6]{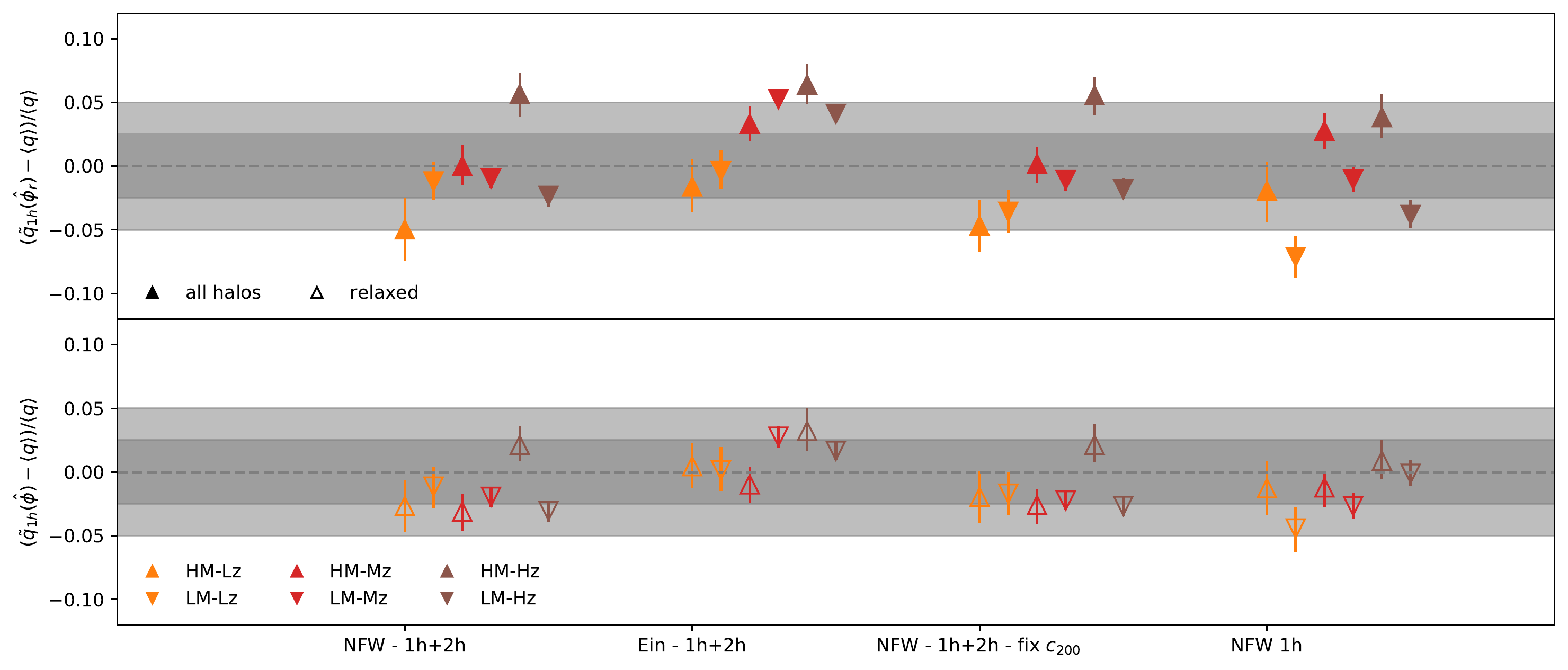}
    \caption{Comparison between the lensing shape estimates of the main halo component $\tilde{q}_{1h}$ and the mean of the semi-axis ratios obtained using the standard inertia tensor considering the 2D dark-matter particle locations, $\langle q \rangle$. In the x-axis we specify the adopted models used to fit the lensing profiles, described in \ref{subsec:biasmodel}. The colour code and symbols represent each sub-sample of halos as defined in Table \ref{tab:sampdef}. In the upper (bottom) panel, lensing estimates are obtained using the standard (reduced) orientation to compute the quadrupole profiles and we consider all (only relaxed) halos within the mass and redshift bins of the sub-samples to compute the average shapes.}
    \label{fig:model}
\end{figure*}

We also obtain that when computing quadrupole profiles using $\hat{\phi}$, lensing shapes are more in agreement with the mean of the semi-axis ratios that include only relaxed halos. On the other hand, when computing the quadrupole components using the reduced direction, $\hat{\phi}_r$, lensing estimates are more in agreement with the mean standard shapes of the whole halo sample, with a larger dispersion for the higher mass sub-samples. Therefore, although reduced estimates predict rounder shapes, the orientation given by the reduced tensor adequately trace the main direction of the mass distribution.

A possible physical interpretation of these results could be the correlation between the halo substructure and the chosen orientation to perform the stacking.
Since the lensing analysis is sensitive to the smoothed mean surface mass distribution, the effect of the substructure vanishes. Therefore, it can be indicating that $\hat{\phi}_r$ direction is more correlated to the halo substructure since it is related to the mean of the whole halo sample regardless of its relaxation state. However, this can be also indicating a non-linear relation between the shape lensing estimate, $\tilde{q}_{1h}$, and the mean of the semi-axis ratios of the particle distribution, $\langle q \rangle$, in which biases tend to be compensated due to the halo selection.


\section{Evaluation of potential bias effects}
\label{sec:bias}

In this section we evaluate the impact of potential biases introduced when using observational data to measure halo shapes using weak-lensing stacking techniques. In particular, we consider two main sources of bias, one related to the adopted modelling and the other related to the fact that the halo centre and projected main orientations are unknown and some assumptions need to be made in order to apply the stacking procedure. 

\subsection{Modelling induced systematics}
\label{subsec:biasmodel}

We evaluate the impact of the modelling in fitting the semi-axis ratios by considering three alternative approaches to the one presented in \ref{subsec:mono}. We refer to this described model in the Figures as \textsf{NFW - 1h+2h}. 

The first alternative model considers a Einasto density profile instead of the NFW, which is defined as \citep{Einasto1989,Retana-Montenegro2012}:
\begin{equation} \label{eq:ein}
\rho(r) = \rho_{-2} \exp{\left( -\frac{2}{\alpha} \left[ \left( \frac{r}{r_{-2}} \right) ^\alpha  - 1 \right] \right)},
\end{equation}
where $\rho_{-2}$ sets the amplitude of the profile and is the density at $r = r_{-2}$, i.e., at the radius where the logarithmic slope of the density profile is $-2$ and which is the equivalent to the NFW scale-length, $r_s$. This model is characterised by a logarithmic slope which is a power law function of the radius and it is set in terms of the extra parameter, $\alpha$. 
The concentration of the profile is characterised by $c_{200} = r_{200}/r_{-2}$. According to previous studies, the Einasto model provides a more precise description of the radial density distribution in dark matter halos, which is also more robust against the variations in fitting details as, for example, the radial range used to fit the parameters \citep{Gao2008,Meneghetti2014,Gonzalez2022}. By considering the projection of the 3D profile defined in Eq. \ref{eq:ein}, 
we model the radial surface density introducing this model as $\Sigma_{1h}$ in Eq. \ref{eq:smodel}. We refer to this model in the Figures as \textsf{Ein - 1h+2h}. In this case we optimise the log-likelihood (Eq. \ref{eq:loglmono}) including the three parameters that characterise the model, $\ln{\mathcal{L}}(\Delta \Sigma | M_{200},c_{200},\alpha)$. We fit $\alpha$ considering a flat prior, $0.2 < \alpha < 0.45$.

The second approach we adopt to evaluate the impact of modelling is to fix the NFW concentration in the analysis by using the concentration relations with mass and redshift presented by \citet{Diemer2019}. Since there is a well-known mass-concentration relation, the lack of information in the inner regions of the density profiles could lead to biased results. Thus, and approach commonly adopted in observational studies is to fix the concentration in the weak-lensing analyses \citep{Uitert2012,Kettula2015,Pereira2018,Gonzalez2021a}. This model is labelled as \textsf{NFW - 1h+2h - fix} $c_{200}$. Taking this into account we optimise Eq. \ref{eq:loglmono} varying only the mass, $\ln{\mathcal{L}}(\Delta \Sigma | M_{200})$.

Then, we alternatively decide to restrict the fitted radial range for the high- and low-massive sub-samples up to $2 h^{-1}$\,Mpc and $1 h^{-1}$\,Mpc, respectively, in order to avoid including the 2-halo term in the modelling (\textsf{NFW - 1h}). Limiting radial ranges are selected roughly at the radial distance in which the contribution to the tangential quadrupole component profile of the 2-halo term is higher than the one related to the main halo (comparison between green and orange lines in Fig. \ref{fig:profile}). For this adopted model we only fit $q_{1h}$ from the quadrupole profiles.

To illustrate the fitting performance of each adopted model, we show derived profiles and the fitted tested models in Fig \ref{fig:promodel} for the HM-Lz considering all the halos in the adopted redshift and mass bin. It can be noticed that the largest differences are obtained when fitting only the main halo component. In Fig. \ref{fig:model} we show the scaled differences between the lensing and the particle distribution shapes. Taking into account the results presented in 
Sec. \ref{subsec:comp}, we compare $\tilde{q}_{1h}(\hat{\phi})$ and $\tilde{q}_{1h}(\hat{\phi}_r)$ estimates with the mean shapes of the relaxed and the whole sub-samples of halos, respectively. We first highlight that regardless of the adopted modelling, shapes can be accurately constrained to within $5\%$. The better agreement in shape estimates is accomplished when using the standard orientation to compute the quadrupole profiles. However, to compare with the particle distribution it is necessary to discard non-relaxed halos, thus it depends on this selection.

\subsection{Bias introduced by using luminous tracers}

In order to apply the stacking techniques when using observational data, there are some assumptions needed regarding the total surface density distribution which are not known a priori, such as the density peak and its main orientation. First, we need to adopt a given centre to compute the profiles. A commonly adopted approach is to fix the cluster centre at the location of the BCG. However, it is well known that in merging galaxy clusters there are significant offsets between the host halo and the galaxy position, which may bias the lensing profiles at the inner regions. 

Secondly, in order to compute the quadrupole profiles we need to set the position angle of the main direction of the surface distribution. Until now, we adopted the direction estimates given by the dark matter particle positions, computed using the inertia tensors. In observational studies several observable tracers have been proposed in order to estimate the main orientation of the clusters, such as the BCG \citep{Uitert2017,dong2014}, the intra-cluster light \citep{Montes2019,Alonso-Asensio2020,Sampaio-Santos2021,Gonzalez2021} and the locations of the galaxy members \citep{Paz2006,shin2018,Uitert2017}. Regardless of the adopted tracer, there is a bias introduced due to the misalignment with respect to the dark-matter particle distribution that will impact only on the quadrupole profile components.

\begin{figure*}
    \centering
    \includegraphics[scale=0.6]{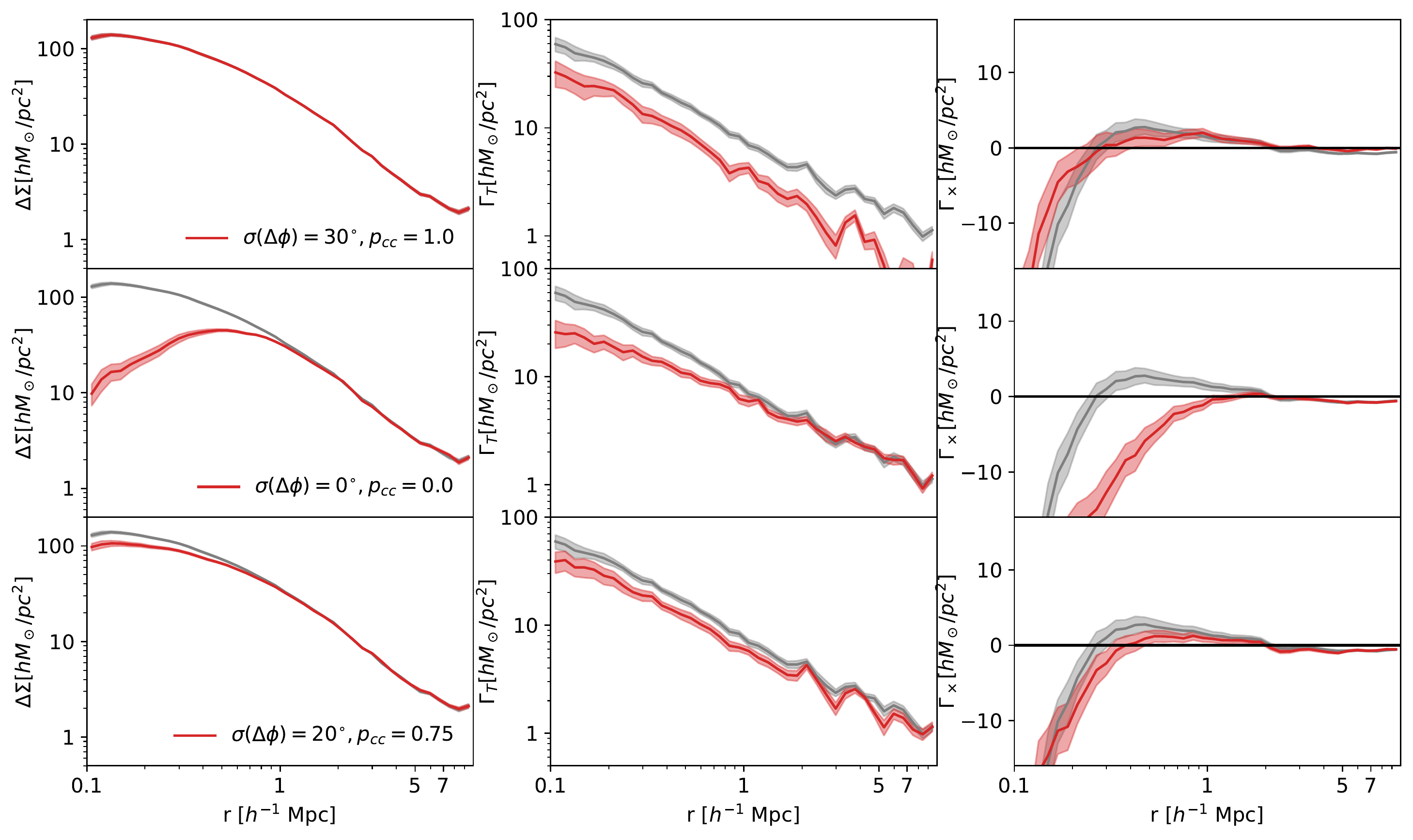}
    \caption{Density contrast (first column), tangential (second column) and cross quadrupole (third column) profiles obtained for the stacked HM-Lz sub-sample of halos in grey. Grey profiles are computed without an introduced miscentering and misalignment. In red we show the profiles when a misalignment distribution with $\sigma(\Delta \phi) = 30^\circ$ is introduced to compute the quadrupole components (first row), when the miscentering is  is introduced for the whole sample of clusters (second row) and, finally, (third row) a more realistic situation, when $25 \%$ of the stacked halos are miscentered and 
    misaligned considering a distribution with $\sigma(\Delta \phi) = 20^\circ$.}
    \label{fig:mis}
\end{figure*}

To illustrate the impact on the profiles due to miscentering and misalignment, we show in Fig. \ref{fig:mis} the computed profiles for the HM-Lz sub-sample when these effects are introduced. In this section we inspect the impact of the mentioned biases in recovering the halo shapes.

\subsubsection{Impact of misalignment}

Projected cluster ellipticity measurements using weak-lensing stacking techniques are affected by the misalignment $\Delta \phi$ of each cluster, defined as the angle between the orientation of the dark matter particle distribution and the estimated angle used to compute the quadrupole component profiles.
Therefore, the lensing signal is diluted by a factor given by the mean of $\cos(2\Delta \phi)$ for the sample of clusters considered \citep{Clampitt2016,vanUitert2017}.  Hence, the measured lensing ellipticity will be biased to lower values compared to the projected dark matter distribution:
\begin{equation} 
\label{eq:misal}
    \epsilon_{1h} = \langle \cos(2 \Delta \phi) \rangle \times \epsilon = D \times \epsilon,
\end{equation}
where $\epsilon$ is the projected ellipticity of the total dark matter distribution ($\epsilon = (1-q)/(1+q)$) and $D$ is the dilution factor. Since the true misalignment for each cluster is unknown in observational studies, $D$ can be estimated by assuming a Gaussian distribution for the misalignments with a given standard deviation, $\sigma(\Delta \phi)$, computed with mock realisations: 
\begin{equation}
    D = \int_0^{2\pi} G(\sigma(\Delta \phi)) \cos(2 \Delta \phi) d\Delta \phi.
\end{equation}
According to hydrodinamical simulations the expected dilution introduced when adopting luminous tracers to estimate the main semi-axis direction is roughly within $0.6$ and $0.9$, depending on the adopted tracer, the cluster mass and relaxation state \citep{Gonzalez2021}. This corresponds to a standard deviation within $10^\circ$ and $30^\circ$.

To inspect the impact of misalignment in the halo shape measurements, we randomly introduce a $\Delta \phi$ added to the known main directions $\hat{\phi}$ and $\hat{\phi}_r$. The value of $\Delta \phi$ for each halo is selected from a Gaussian distribution with standard deviations of $10^\circ$, $20^\circ$ and $30^\circ$. This introduced misalignment will only impact the quadrupole components (upper panel of Fig. \ref{fig:mis}), thus we fit only the semi-axis ratios from these misaligned quadrupole profiles. 

Ellipticities from the fitted semi-axis ratios scaled by the measured ones without misalignment and according to the projected dark matter ellipticity, are shown in Table \ref{tab:misal}. In general, fitted ellipticities are lowered according to the expected dilution. Even $\epsilon_{2h}$ can be well recovered when considering $D$, although with a larger uncertainty due to a higher dispersion around the dilution value.

\begin{table*} 
 \caption{Results considering misalignment and miscentering}
    \centering
    \begin{tabular}{c c c c c c c}
    \hline
    \hline
Sub-sample name  &  \multicolumn{3}{c}{standard} & \multicolumn{3}{c}{reduced} \\
             & $\tilde{\epsilon}^{mis}_{1h} / \tilde{\epsilon_{1h}}$ 
             & $\tilde{\epsilon}^{mis}_{2h} / \tilde{\epsilon_{2h}}$ 
             & $\tilde{\epsilon}^{mis}_{1h} / \epsilon$ 
             & $\tilde{\epsilon}^{mis}_{1h} / \tilde{\epsilon_{1h}}$ 
             & $\tilde{\epsilon}^{mis}_{2h} / \tilde{\epsilon_{2h}}$ 
             & $\tilde{\epsilon}^{mis}_{1h} /  \epsilon $ \\
\hline
\multicolumn{7}{c}{$p_{cc} = 1, \sigma(\Delta \phi) = 10^{\circ} - D \sim 0.94$} \\
\hline
HM-Lz & $0.94$ & $1.13$ & $0.99$ & $0.93$ & $0.98$ & $1.01$ \\ 
LM-Lz & $0.95$ & $0.92$ & $0.98$ & $0.96$ & $0.85$ & $0.98$ \\ 
HM-Mz & $0.93$ & $0.90$ & $0.99$ & $0.94$ & $0.86$ & $0.94$ \\ 
LM-Mz & $0.93$ & $0.82$ & $0.97$ & $0.94$ & $0.92$ & $0.96$ \\ 
HM-Hz & $0.91$ & $0.93$ & $0.87$ & $0.91$ & $0.89$ & $0.83$ \\ 
LM-Hz & $0.99$ & $0.88$ & $1.06$ & $0.99$ & $0.91$ & $1.03$ \\ 
\hline
\multicolumn{7}{c}{$p_{cc} = 1, \sigma(\Delta \phi) = 20^{\circ} - D \sim 0.78$} \\
\hline
HM-Lz & $0.82$ & $0.87$ & $0.87$ & $0.77$ & $0.93$ & $0.84$ \\ 
LM-Lz & $0.81$ & $0.73$ & $0.83$ & $0.81$ & $0.73$ & $0.83$ \\ 
HM-Mz & $0.82$ & $0.81$ & $0.88$ & $0.81$ & $0.83$ & $0.81$ \\ 
LM-Mz & $0.76$ & $0.76$ & $0.79$ & $0.78$ & $0.81$ & $0.79$ \\ 
HM-Hz & $0.78$ & $0.97$ & $0.74$ & $0.83$ & $0.87$ & $0.75$ \\ 
LM-Hz & $0.82$ & $0.83$ & $0.87$ & $0.83$ & $0.93$ & $0.87$ \\ 
\hline
 \multicolumn{7}{c}{$p_{cc} = 1, \sigma(\Delta \phi) = 30^{\circ} - D \sim 0.58$} \\
 \hline
HM-Lz & $0.59$ & $0.44$ & $0.62$ & $0.54$ & $0.26$ & $0.59$ \\ 
LM-Lz & $0.60$ & $0.58$ & $0.62$ & $0.63$ & $0.49$ & $0.64$ \\ 
HM-Mz & $0.54$ & $0.67$ & $0.58$ & $0.58$ & $0.69$ & $0.58$ \\ 
LM-Mz & $0.51$ & $0.54$ & $0.53$ & $0.54$ & $0.62$ & $0.55$ \\ 
HM-Hz & $0.56$ & $0.62$ & $0.53$ & $0.57$ & $0.56$ & $0.52$ \\ 
LM-Hz & $0.52$ & $0.39$ & $0.55$ & $0.51$ & $0.45$ & $0.53$ \\ 
\hline
 \multicolumn{7}{c}{$p_{cc} = 0.75, \sigma(\Delta \phi) = 0^{\circ} - D = 1.0$} \\
 \hline
HM-Lz & $0.78$ & $1.17$ & $0.82$ & $0.78$ & $1.14$ & $0.85$ \\ 
LM-Lz & $0.81$ & $1.10$ & $0.84$ & $0.81$ & $1.05$ & $0.83$ \\ 
HM-Mz & $0.88$ & $1.01$ & $0.94$ & $0.86$ & $1.05$ & $0.86$ \\ 
LM-Mz & $0.79$ & $0.96$ & $0.82$ & $0.81$ & $1.06$ & $0.83$ \\ 
HM-Hz & $0.86$ & $1.04$ & $0.82$ & $0.84$ & $1.02$ & $0.76$ \\ 
LM-Hz & $0.84$ & $0.98$ & $0.90$ & $0.89$ & $0.98$ & $0.93$ \\ 
\hline
\multicolumn{7}{c}{$p_{cc} = 0.75, \sigma(\Delta \phi) = 20^{\circ} - D \sim 0.78$} \\
\hline
HM-Lz & $0.67$ & $0.86$ & $0.71$ & $0.67$ & $0.87$ & $0.73$ \\ 
LM-Lz & $0.61$ & $0.95$ & $0.63$ & $0.66$ & $0.96$ & $0.68$ \\ 
HM-Mz & $0.68$ & $0.73$ & $0.73$ & $0.70$ & $0.86$ & $0.70$ \\ 
LM-Mz & $0.71$ & $0.97$ & $0.74$ & $0.68$ & $0.98$ & $0.69$ \\ 
HM-Hz & $0.70$ & $0.84$ & $0.67$ & $0.71$ & $0.76$ & $0.65$ \\ 
LM-Hz & $0.79$ & $0.62$ & $0.84$ & $0.63$ & $0.73$ & $0.66$ \\ 
\hline
\end{tabular}
\begin{flushleft}
Columns: (1) Name of the sub-sample of halos as described in Table \ref{tab:sampdef} (2) Lensing elongation of the main halo component, obtained using the standard orientation, considering miscentering and misalignment scaled by the unbiased lensing estimate; (3) Lensing elongation of the main halo component, obtained using the standard orientation, considering miscentering and misalignment scaled by the unbiased lensing estimate; (4) Lensing elongation of the neighbouring distribution, obtained using the standard orientation, considering miscentering and misalignment scaled by the projected dark-matter particle elongation obtained from the mean of the 2D semi-axis ratio computed according to the standard inertia tensor. (5), (6) and (7) are analogous but lensing estimates are obtained using the reduced orientations to compute the quadrupole profiles. In this case, we include only relaxed halos to compute $\epsilon$. The different sets of rows indicate the introduced miscentering and misalignment from which $\tilde{\epsilon}^{mis}$  are obtained.
\end{flushleft}
    \label{tab:misal}
\end{table*}

\subsubsection{Impact of miscentered halos}

A possible 
offset between the centre of the halo host and the location of the adopted centre to compute the profiles 
will affect derived profiles by flattening the lensing signal at the inner regions. 
We model the 
miscentering of the halos following the formalism 
proposed by \citet{Yang2006,Johnston2007,Ford2014}, which considers two cluster populations. The first includes well-centred clusters for which an offset lower than $\sim 50$\,kpc it is expected, and that constitutes a fraction $p_{cc}$ of the total sample. The second population of the stacked galaxy clusters includes merging systems that are highly miscentered. The adopted centres of these clusters are expected to be shifted from the halo density peak by a radial offset $r_\text{mis}$, which follows a Gamma distribution \citep{Zhang2019,Yan2020},
\begin{equation}
\label{eq:Pdist}
    P(r_\text{mis}) = \frac{r_\text{mis}}{(\tau R_{200})^2} \exp{\left(-\frac{r_{s}}{\tau R_{200}} \right)},
\end{equation}
where $\tau R_{200}$ is the dispersion of the distribution, which is scaled according to the  $R_{200}$. Taking this distribution into account we randomly introduce an $r_\text{mis}$ offset to a $(1 - p_{cc})$ fraction of halos. We set $p_{cc} = 0.75$ and $\tau = 0.2$ according to previously reported observational works \citep{Simet2017,McClintock2019} based on the redMaPPer cluster sample \citep{Rykoff2014,Rykoff2016}. 

In the middle panel of Fig \ref{fig:mis} we show how the miscentering affects the profiles when the whole sample of halos is considered to follow the offset distribution, i.e. $p_{cc} = 0$, to highlight the effect. 
As we can see, 
it affects significantly at the inner radial ranges of all the computed profiles. Although this bias affects the density contrast profile, thus it will impact the halo mass and concentration estimates, we fit only the quadrupole components using the previously fitted results from the $\widetilde{\Delta \Sigma}$ profiles without miscentering. In that way we assume that these parameters can be well recovered by properly modelling the miscentered $\widetilde{\Delta \Sigma}$ component, as proposed in \citet{Yang2006,Johnston2007,Ford2014}. Thus, we only the evaluate the impact of this effect in the quadrupole components.

In Table \ref{tab:misal} we show the measured ellipticities when the miscentering is considered. In general, projected ellipticities related to the main halo component are biased by roughly a $20 \%$, while the elongation related to the neighbouring component is less biased, given that this contributes mainly at the outer regions of the profiles which are less affected by the miscentering effect. 

\subsection{Combined effects}

In order to consider a realistic situation, we compute the profiles for the whole sub-samples of halos introducing a misalignment according to a Gaussian distribution with a standard deviation $\sigma(\Delta \phi) = 20^\circ$, as well as introducing a radial miscentering following Eq. \ref{eq:gamma} to $25 \%$ of the halos ($p_{cc} = 0.75$). We fit the quadrupole profiles using all the models tested presented in \ref{subsec:biasmodel}. Computed scaled lensing elongations are shown in Table \ref{tab:misal}, obtaining biased lensing estimates by $\sim 30\%$.

Once the dilution due to misalignment is considered, unbiased shape estimates can be recovered by modelling the impact of miscentering. However, since the effects are combined, it is not straightforward how they can be disentangled. Moreover, the miscentering is not easy to model since it can be also coupled to the main direction of the mass distribution. Another strategy could be to select the clusters for the analysis in order to exclude those with expected significant offsets or with clear signs of being merging systems. We discuss this strategy in the next section.

\section{Proposed observational strategy}
\label{sec:strategy}

\begin{figure*}
    \centering
    \includegraphics[scale=0.6]{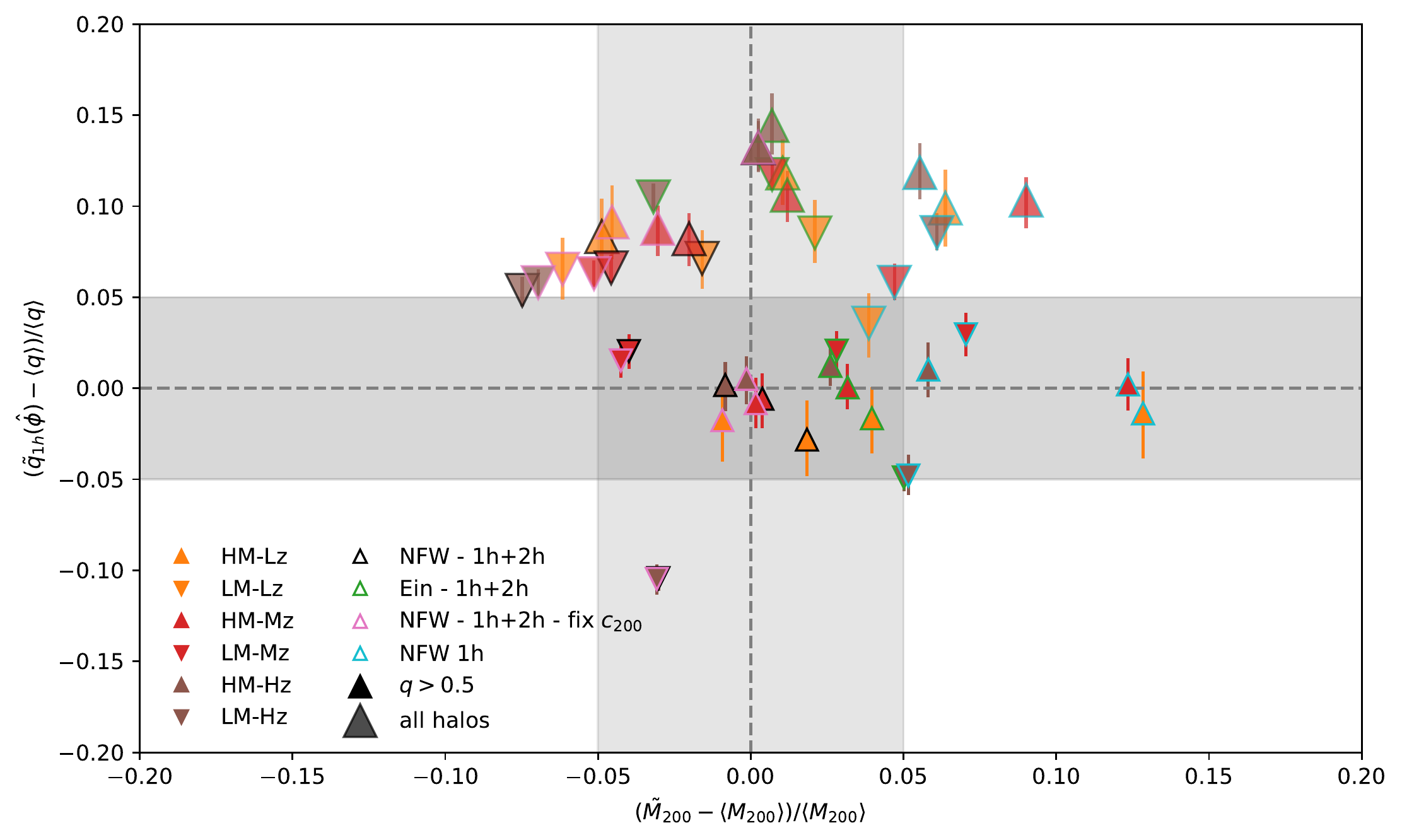}
    \caption{Comparison between lensing shape estimates of the main halo component, $\tilde{q}_{1h}$, and the mean of the 2D semi-axis ratio obtained using the standard inertia tensor, $\langle q \rangle$, vs. the comparison of the lensing $\tilde{M}_{200}$ masses with the mean of the fitted 2D particle density profiles, $\langle M_{200} \rangle$. Lensing estimates are obtained using the standard orientation to compute the quadrupole profiles and we consider the same stacked halos to compute the average shapes and 2D masses. The colour code and symbols represent each sub-sample of halos as defined in Table \ref{tab:sampdef} and the border of the triangles refers to each adopted model to compute lensing estimates. Shaded and larger symbols are related to the results using the whole sample of halos within the redshift and mass ranges, while shaded and smaller symbols are related to the estimates that only include rounder halos ($q > 0.5$), to discard those highly un-relaxed.}
    \label{fig:allbias}
\end{figure*}

\begin{figure*}
    \centering
    \includegraphics[scale=0.6]{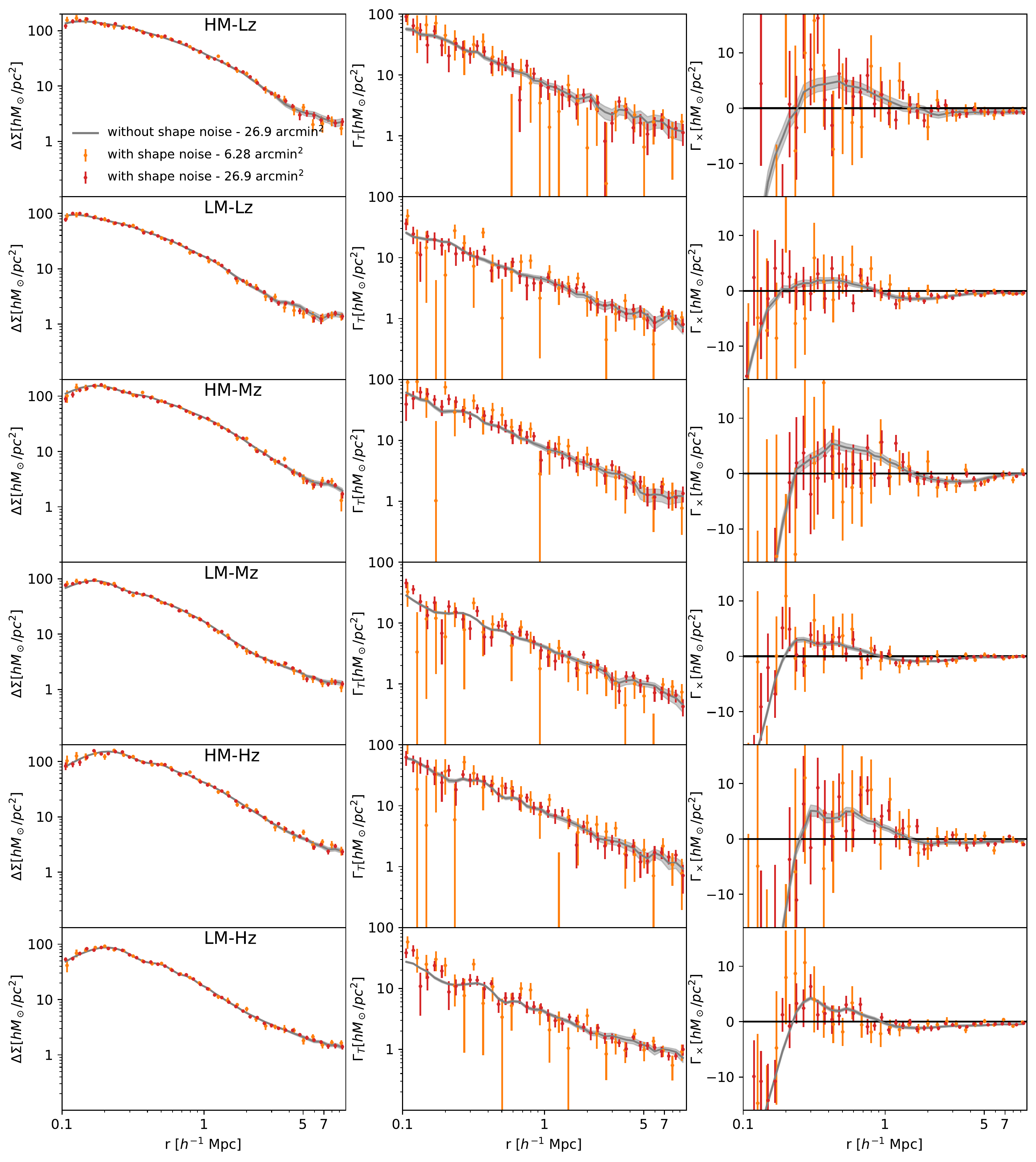}
    \caption{Stacked density contrast profiles ($\Delta \Sigma$, first column), tangential ($\Gamma_{\text{T}}$, second column) and cross ($\Gamma_{\times}$, third column) quadrupole component profiles for the halos with $q>0.5$ in the sub-samples defined in Table \ref{tab:sampdef}. Quadrupole profiles are computed considering the standard orientation ($\hat{\phi}$). Grey solid lines correspond to the profiles computed using the full sample of source galaxies by combining the \textit{shear} components without accounting for shape-noise. Red and orange dots are the computed profiles considering shape-noise using the full sample of source galaxies and a sub-set randomly selected to match DES density. Errors are computed according to the square root of the diagonal components of the covariance matrix.} 
    \label{fig:profile_withnoise}
\end{figure*}

\begin{figure*}
    \centering
    \includegraphics[scale=0.6]{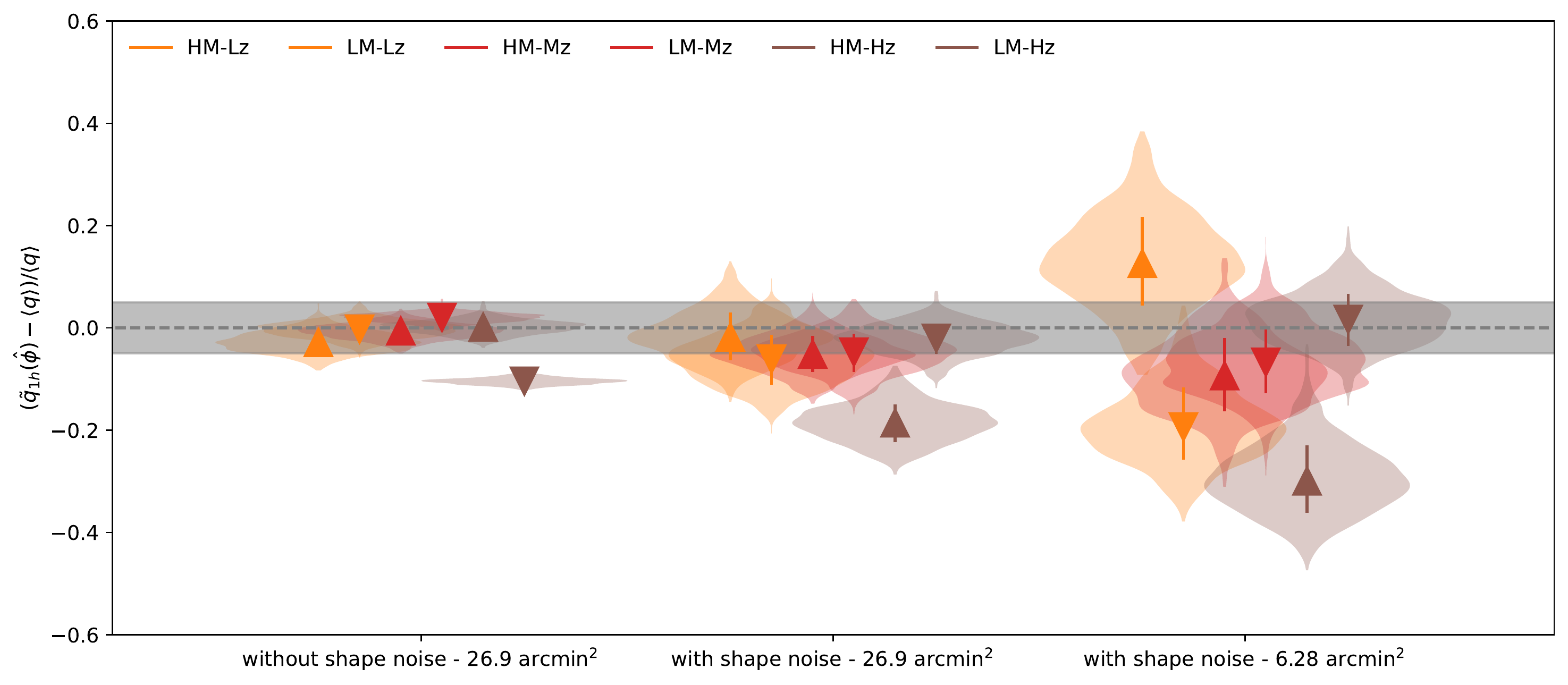}
    \caption{Comparison between the lensing shape estimates of the main halo component $\tilde{q}_{1h}$ and the mean of the semi-axis ratios obtained using the standard inertia tensor considering the 2D dark-matter particle locations, $\langle q \rangle$. In the x-axis we specify the source sample density used to compute the profiles and we indicate if shape-noise is included. The colour code and symbols represent each sub-sample of halos for halos with $q>0.5$ specified in Table \ref{tab:sampdef}. Violin plots correspond to the scaled fitted posterior density distributions.}
    \label{fig:reswithnoise}
\end{figure*}

\subsection{Discarding highly un-relaxed systems}

In order to mitigate the impact of miscentering and to derive less biased shape estimates, we propose to discard highly elongated halos from the stacked analysis, since these halos are more probably undergoing a merger (see Fig. \ref{fig:qrelax}). Using observed clusters, this selection can be performed by considering only those clusters with a rounder galaxy member distribution. This selection strategy can be tested in a posterior work that includes the galaxy member distribution to estimate the cluster orientation. Since there is a expected relation between the halo elongation and the relaxation state according to the offset centre (Fig. \ref{fig:qrelax}), another approach could be selecting those clusters with BCGs that have particular photometric properties as it is done for the redMaPPer cluster catalogue, in which a centre probability is provided. 

It is important to take into account that the proposed halo selection to perform the analysis will, in principle, introduce a mass bias, due to the interplay between halo masses and shapes. In fact, we obtain that, when halos with $q<0.5$ are discarded from the analyses, mean masses for the considered sub-samples are roughly $7\%$ and $5\%$ lower, for the higher (HM) and lower mass halos (LM), respectively. It is therefore important to perform equivalent cuts when comparing predicted shapes with observable parameters.

We perform the same weak-lensing analysis with the same defined sub-samples, but discarding those halos with $q < 0.5$. The number of stacked halos is specified in Table \ref{tab:sampdef}. Mean shapes for halos with $q > 0.5$, do not depend on the relaxation state, with $\langle q \rangle$ in agreement within $< 1\%$ when discarding un-relaxed systems. The advantage of this analysis, besides mitigating miscentering effects, is that it is not necessary to consider a relaxation criteria to compare the samples since highly un-relaxed halos are already discarded. When considering the reduced orientation to compute the quadrupoles, lensing shapes are more elongated than the dark-matter distribution, with $(\tilde{q}_{1h} - \langle q \rangle) / \langle q \rangle \sim -0.1$ for all the halo sub-samples. On the other hand, when considering the standard orientation shapes are in excellent agreement within $\sim 5 \%$.

Fitted masses and semi-axis ratios, using the standard orientation, compared with the mean values derived according to the dark-matter particle distribution are shown in Fig. \ref{fig:allbias}. In this plot we show the lensing shape estimates when considering the whole sub-sample of halos and when highly elongated halos are discarded, keeping only those with $q > 0.5$. In the first case, lensing semi-axis ratios are biased and predict rounder shapes, notice that we are not discarding un-relaxed systems as in the bottom panel of Fig. \ref{fig:model}. On the other hand, when considering rounder halos for the analysis, shapes can be properly recovered. This result indicates that lensing shapes of un-relaxed halos are mainly rounder than those obtained according to the particle distribution, since the presence of substructure is mitigated in the stacking procedure. 

Masses are compared with the mean values from the fitted 2D density profiles, considering the whole sample of halos. We decide to use the mean fitted 2D masses instead of the one derived from the 3D distributions, to account for the projection effects that can bias the results \citep{Debackere2022}. As 
we can see, masses show a large scatter when comparing the mean estimates. As mentioned, the observed differences can be related to the fitting radial range, as well as deficiencies in the modelling \citep{Meneghetti2014,Child2018,Debackere2022,Gonzalez2022}. However, regardless of the biases introduced in the mass measurements, projected semi-axis ratios can be properly well recovered using the lensing analysis.

\subsection{\textbf{Predicted results including shape noise}}

Until now, all the estimated values were obtained by using the \textit{shear} components without accounting for the introduced shape noise, due to the intrinsic shapes of the galaxies in the source sample. To test how the semi-axis ratio of cluster mass distributions can be recovered using observational data-sets, we consider this noise by adding intrinsic ellipticity components, ($e_1^s$,$e_2^s$), to the \textit{shear} quantities, ($\gamma_1^s$,$\gamma_2^s$), given in MICEv2.0 mock catalogue. 

The intrinsic ellipticity components are computed as described in
\citet{Hoffmann2022}. Galaxies are thereby modelled as 3D ellipsoids,
whose axis ratios depend on galaxy magnitude, colour and redshift, which we obtain from the galaxy mock catalogue used to select source galaxies (see \ref{subsec:sources}).
The model has been calibrated to reproduce distributions of 2D axis ratios
from COSMOS observations.
We neglect the effect of intrinsic alignments given that it is expected to be negligible due to the large redshift bin in which source galaxies are combined in the stacking procedure.  

We compute the profiles according to the defined estimators by considering the predicted observed ellipticity components:
\begin{equation}
    e_1 = \gamma_1 + e_1^s, \\
    e_2 = \gamma_2 + e_2^s,
\end{equation}
instead of the \textit{shear} in Eqs. \ref{eq:rprof}--\ref{eq:profgammacross}. The sample of stacked halos is the same proposed in the above subsection, with $q>0.5$. In order to compute the profiles we consider the full source sub-sample and a randomly selected sub-set to reproduce the same galaxy density as for DES. In this case we compute the profile using 30 instead of 40 radial bins with the same radial range as described in Sec. \ref{subsec:estimators}. Predicted observed profiles are shown in Fig. \ref{fig:profile_withnoise}. We fit quadrupole profiles fixing the masses and concentrations derived from the fitted $\Delta \Sigma$ without considering shape noise, since, due to the larger signal-to-noise ratio of these profiles, we do not expect significant differences in these parameters. 

In Fig. \ref{fig:reswithnoise} we show the results for the lensing fits compared with the mean $q$ of stacked halos. In this plot we show the obtained posterior probability distributions of the fitted semi-axis ratios using the full source galaxy sample with and without considering shape-noise and for a lower source density sample considering shape-noise. Albeit with a low precision, 
the results show that shapes can be properly constrained with the current DES survey, while a high precision could be achieved with future wide field weak-lensing data sets.

\section{Summary and conclusions}
\label{sec:conclusion}

Projected mean halo shapes, albeit with a large scatter, can be related to their total mean mass, which is driven by the assembly process that generates the density distribution. Taking this into account, 
the projected elongation of the mean surface density distribution of halo samples with different masses and redshifts, can be used to test the halo formation and evolution scenario, thus the hierarchical accretion process. Beyond that, it can also set constraints to the nature of dark-matter constituents. Furthermore, characterising the mean projected shapes of these systems allows one to quantify the biases introduced when using these halos to set constraints on cosmological parameters, such as in cluster cosmology, since the halo triaxiality is expected to be one of the main sources of systematics in these studies \citep{Herbonnet2019,Abbott2020,Zhang2022}. 

Given the several potential interests of obtaining reliable cluster shapes, we have performed a weak-lensing analysis applied to several samples of halos in MICE-GC numerical simulation. In order to perform this study, we use halo shapes obtained from the projected distribution of the dark-matter particles through the inertia tensor. We consider two tensor definitions, commonly used in the literature, to obtain the shapes, the standard and the reduced inertia tensor. From these tensors we obtain the 2D semi-axis ratio and the position angle of the main elongation direction that characterise the projected distribution. To perform the lensing study we consider the main halo projected orientation and compute the quadrupole profiles of the surface density distribution, related to the mean halo elongation. We fit the lensing profiles considering an elliptical NFW model plus an elongated 2-halo term to account for the neighbouring mass component. The semi-axis ratios derived from the inertia tensors were contrasted with the lensing aligned elongation component, allowing for a direct comparison between both parameters. 

Our results show that, when adopting the halo orientations given by the standard and reduced tensors, fitted lensing shapes are in agreement with the standard semi-axis ratio related with the projected particle distribution. In particular, lensing shape estimates computed taking into account the orientations given by the reduced tensor, are in agreement with the mean of the standard semi-axis ratio for the same sample of halos considered in the lensing analysis. On the other hand, when considering the standard orientation, lensing estimates are in agreement with the mean of the semi-axis ratio obtained after discarding non-relaxed halos. This might indicate that substructure has a tighter correlation with the reduced-inferred direction. We obtain that this result is model independent by analysing three different alternative approaches in the fitting procedure of the density profiles. 

We have also tested the impact in the shape recovering  of two main sources of systematics introduced in observational studies. One of the main introduced bias comes from the fact that the main direction of the surface density distribution is not known a priori. Thus, luminous tracers associated to the galaxy member distribution are commonly used to estimate the main elongation direction. This introduces a bias in the lensing study due to the possible misalignment between the adopted tracer and the underlying mass distribution. We have tested the effect of this misalignment obtaining that the projected lensing ellipticities can be properly recovered when the misalignment distribution width is known. 

Another effect introduced when using observational data is related to the inferred halo centres. This is commonly referred to as miscentering and is due to the offset between the maximum surface density distribution and the adopted centres to compute the profiles. We have tested the influence of this effect which mainly impact on the inner radial ranges of the computed profiles, predicting rounder shapes since it lowers the lensing signal. Assuming that $25\%$ of the clusters that are considered in the analysis are undergoing a merger and thus, present larger offsets, we obtain that without accounting for this systematic, lensing elongation estimates can be biased up to a $\sim 20\%$. 

In order to mitigate the miscentering effect and to properly recover the shape parameters with the lensing analysis, we propose as strategy to discard those clusters that are undergoing a merger and show a highly elongated dark-matter particle distribution. Once these halos are discarded from the analysis, we constrain the shapes to within $5\%$ by using the standard orientation to perform the stacking study. This approach has also the advantage of simultaneously discarding highly un-relaxed systems, simplifying the comparison between the shapes derived from lensing and the dark-matter particle distribution. In general shapes described according to the lensing estimate for un-relaxed halos are rounder since the analysis is centred on the densest halo component and substructure effects are blurred due to the stacking. On the contrary, these halos show a highly elongated particle distribution.  

Finally, we also include in the analysis the noise introduced due to the intrinsic shape of the galaxies in the source sample and predict the accuracy of the shape determinations when using wide-field \textit{shear} catalogues. We obtain that a high accuracy can be achieved using the data of future surveys such as LSST and Euclid.

In summary, the present analysis tests weak-lensing stacking techniques as a tool that allows one measure the main projected shapes of a sample of halos. By applying these procedures we derive extra parameters besides those usually obtained in current analysis, i.e. mass and concentrations, further characterising the density distribution of halos. 
Our results show that the measured shapes can be well compared with the one expected from numerical simulations and are almost unaffected by the choice of the model. Furthermore, the techniques detailed in this paper allow us to test the presence of halo substructure and the elongation of the neighbouring mass distribution as an extra information that can be modelled and provided by the analysis.

\section*{Acknowledgements}

We kindly thank the anonymous referee for her/his careful reading and comments that significantly improve this work. 
This project has received funding from the European Union’s Horizon 2020 Research and Innovation Programme under the Marie Sklodowska-Curie grant agreement No 734374. This work was also partially supported by Agencia Nacional de Promoción Científica y Tecnológica (PICT-2020-SERIEA-01404), the Consejo Nacional de Investigaciones Científicas y Técnicas (CONICET, Argentina)
and the Secretaría de Ciencia y Tecnología de la Universidad Nacional de Córdoba (SeCyT-UNC, Argentina). This work has made use of CosmoHub. CosmoHub has been developed by the Port d’Informació Científica (PIC), maintained through a collaboration of the
Institut de Física d’Altes Energies (IFAE) and the Centro de Investigaciones Energéticas, Medioambientales y Tecnológicas (CIEMAT), and was partially funded by the “Plan Estatal de Investigación Científica y Técnica y de Innovación” program of the Spanish government. MM is partially supported by FAPERJ and CNPq.

\section*{Data Availability}

The data used in this project is available through CosmoHub platform\footnote{\href{https://cosmohub.pic.es/home}{https://cosmohub.pic.es/home}} \citep{Carretero2017,TALLADA2020100391} under the name `MICE HALO properties' and `MICECAT' version 2.



\bibliography{bib}







\bsp	
\label{lastpage}
\end{document}